\documentclass[%
aps,nofootinbib,notitlepage,superscriptaddress,10Limberkipt,prd, twocolumn,
 amsmath,amssymb
]{revtex4-2}

\usepackage[caption=false]{subfig}
\usepackage{braket}
\usepackage{float}
\usepackage{lipsum}
\usepackage{graphicx}
\usepackage{dcolumn}
\usepackage{bm}
\usepackage{natbib}
\usepackage{hyperref}
\hypersetup{
	colorlinks = true,
    linkcolor = Maroon,
    urlcolor  = Maroon,
    citecolor = Maroon
}
\usepackage{afterpage}
\usepackage{placeins}
\usepackage{microtype}
\usepackage{verbatim}
\usepackage[amssymb]{SIunits}
\usepackage{tabularx}
\usepackage[dvipsnames]{xcolor}
\usepackage{tikz}

\usepackage{color}
\definecolor{darkgreen}{RGB}{0,120,0}

\newcommand{\av}[1]{\langle{#1}\rangle}

\newcommand{\vK}{{\bm K}}
\newcommand{\vQ}{{\bm Q}}
\newcommand{\vP}{{\bm P}}
\newcommand{\be}{\begin{equation}}
\newcommand{\ee}{\end{equation}}
\newcommand{\bea}{\begin{eqnarray}}
\newcommand{\eea}{\end{eqnarray}}
\newcommand{\beas}{\begin{eqnarray*}}
\newcommand{\eeas}{\end{eqnarray*}}

\newcommand{\vx}{\mathbf{x}}
\newcommand{\vk}{\mathbf{k}}
\newcommand{\vq}{\mathbf{q}}
\newcommand{\vp}{\mathbf{p}}

\newcommand{\bx}{{\boldsymbol x}}
\newcommand{\by}{{\boldsymbol y}}

\def\gsim{ \lower .75ex \hbox{$\sim$} \llap{\raise .27ex \hbox{$>$}} }
\def\lsim{ \lower .75ex \hbox{$\sim$} \llap{\raise .27ex \hbox{$<$}} }

\def\dalam{\hbox
{\vrule\vbox{\hrule\hbox to 1ex{ \hfill}\kern 1 ex\hrule}\vrule}}

\def\1/2{\hbox{$ {1 \over 2}$ }}

\def\h{\hbar}
\def\i/h{{i \over \h}}

\begin{document}

\title{Massive-ish Particles from Small-ish Scales: Non-Perturbative Techniques for Cosmological Collider Physics from Large-Scale Structure Surveys}


\author{Samuel~Goldstein}
\email{sjg2215@columbia.edu}
\affiliation{Department of Physics, Columbia University, New York, NY 10027, USA}

\author{Oliver~H.\,E.~Philcox}
\affiliation{Department of Physics, Columbia University, New York, NY 10027, USA}
\affiliation{Simons Society of Fellows, Simons Foundation, New York, NY 10010, USA}

\author{J.~Colin~Hill}
\affiliation{Department of Physics, Columbia University, New York, NY 10027, USA}

\author{Lam~Hui}
\affiliation{Department of Physics, Columbia University, New York, NY 10027, USA}

%
\begin{abstract}
\noindent Massive particles produced during inflation impact soft limits of primordial correlators. Searching for these signatures presents an exciting opportunity to uncover the particle spectrum in the inflationary epoch. We present non-perturbative methods to constrain intermediate-mass scalars ($0\leq m/H<3/2$, where $H$ is the inflationary Hubble scale) produced during inflation, which give rise to a power-law scaling in the squeezed primordial bispectrum. Exploiting the large-scale structure consistency relations and the separate universe approach, we derive models for the late-time squeezed matter bispectrum and collapsed matter trispectrum sourced by these fields. To validate our models, we run $N$-body simulations with the ``Cosmological Collider" squeezed bispectrum for two different particle masses. Our models yield unbiased constraints on the amplitude of non-Gaussianity, $f_{\rm NL}^{\Delta}$, from the squeezed bispectrum and collapsed trispectrum deep into the non-linear regime ($k_{\rm max}\approx 2~h/{\rm Mpc}$ at $z=0$). We assess the information content of these summary statistics, emphasizing the importance of sample variance cancellation in the matter sector. We also study the scale-dependent halo bias in our simulations. For mass-selected halos, the non-Gaussian bias estimated from our simulations agrees with predictions based on (i) separate universe simulations and (ii) universal mass functions.  With further work, these results can be used to search for inflationary massive particle production with upcoming galaxy surveys. 
\end{abstract}

\maketitle

\section{Introduction}\label{Sec:Intro}

\noindent In the standard model of cosmology, structures formed from quantum fluctuations that were stretched to macroscopic scales during a period of exponential expansion known as inflation~\cite{Guth:1980zm, Linde:1981mu, Albrecht:1982wi}. Since the Hubble scale during inflation may have been as high as $10^{14}~{\rm GeV}$, understanding the physics responsible for inflation could provide insights into physics beyond the standard model. Remarkably, the statistical properties of the primordial fluctuations encode substantial information about their physical origins. For instance, while the simplest single-field models of inflation produce essentially Gaussian fluctuations~\cite{Maldacena:2002vr, Creminelli:2004yq, Creminelli:2011rh, Pajer:2013ana}\footnote{The simplest, single-field, slow roll inflation does produce some degree of non-Gaussianity from purely gravitational interactions~\cite{Cabass:2016cgp}. However, as far as large-scale structure is concerned, this level of non-Gaussianity can be safely ignored.
}, the presence of extra interactions and additional particles during inflation can introduce non-Gaussianities in the primordial perturbations~\cite{Meerburg:2019qqi, Achucarro:2022qrl}. The search for this primordial non-Gaussianity (PNG) is a central goal of many ongoing and upcoming cosmological surveys~\cite{DESI:2013agm, EUCLID:2011zbd, LSSTDarkEnergyScience:2018jkl, SPHEREx:2014bgr, SimonsObservatory:2018koc, CMB-S4:2016ple}.

Currently, the tightest constraints on various models of PNG come from analyses of the cosmic microwave background (CMB) primary anisotropies~\cite{Planck:2019kim}. While near-term CMB experiments are expected to improve constraints on PNG by roughly a factor of $2-3$ \citep{CMB-S4:2016ple}, large-scale structure (LSS) surveys can, in principle, provide more substantial improvements due to their large effective volume and access to three-dimensional information~\cite{Sailer:2021yzm, Cabass:2022epm}. In practice, extracting PNG from LSS surveys is a messy business because LSS data contain non-Gaussianities due to, \emph{e.g.}, non-linear structure formation, baryonic complications,
and redshift-space distortions. In the quasi-linear regime, perturbative approaches have been used to robustly constrain PNG from galaxy clustering observations~\cite{DAmico:2022gki, Cabass:2022ymb, Cabass:2022wjy, Ivanov:2024hgq, Cabass:2024wob}. Nevertheless, since LSS surveys also provide precision measurements in the non-linear regime, it is valuable to develop techniques to constrain PNG from non-linear LSS observables.

On small scales, non-linearities pose a significant theoretical roadblock towards robust cosmological inference with LSS data. One possibility to circumvent these non-linear nuisances is to exploit the LSS consistency relations~\cite{Kehagias:2013yd, Peloso:2013zw} (see Refs.~\cite{Valageas:2013zda,Creminelli:2013poa,Creminelli:2013mca,Horn:2014rta,Horn:2015dra,Marinucci:2020weg} for more details), which state that, in the absence of PNG and equivalence-principle-violating physics, the ratio of the squeezed bispectrum to the long-wavelength (soft) power spectrum is protected from inverse-powers-of-$q$ poles \cite{Esposito:2019jkb}, \emph{i.e.},
\begin{equation}\label{eq:LSS_CR_statement}
    \lim_{q\ll k}\frac{B(\vq,\vk)}{P(\vq)} \,\,\, \textrm{has no $1/q^\alpha$ poles}.
\end{equation}
Here, $B$ and $P$ are the equal-time, real-space matter bispectrum and power spectrum, respectively
\footnote{In this work, we will assume $B$ and $P$ are computed from the real-space matter field. However, we emphasize that Eq.~\eqref{eq:LSS_CR_statement} is expected to hold for biased tracers of the matter distribution, such as galaxies~\cite{Peloso:2013zw,Kehagias:2013rpa,Creminelli:2013mca,Horn:2014rta,Simonovic:2014yna,Kehagias:2015tda}. Furthermore, the LSS consistency relations take a similar form in redshift space~\cite{Creminelli:2013mca, Creminelli:2013poa,Creminelli:2013nua}.}, $\vq, \vk$ are the momenta (wavevectors) with $q \ll k$ (the squeezed limit), and $\alpha>0$ is a real number. Since Eq.~\eqref{eq:LSS_CR_statement} is a non-perturbative consequence of the symmetries of the LSS equations of motion, it applies for non-linear $k$, outside the perturbative regime. Furthermore, Eq.~\eqref{eq:LSS_CR_statement} can be violated if the primordial density perturbations are not Gaussian. Thus, one can constrain PNG by searching for poles in the \emph{non-linear} squeezed $B/P$ ratio. Indeed, Refs.~\cite{Goldstein:2022hgr, Giri:2023mpg, Goldstein:2023brb} have used this approach to develop estimators for local PNG that are valid in the non-linear regime. In this work, we generalize these estimators to constrain PNG sourced by massive scalar particle exchange during inflation.

Particles with mass $m\approx H$ produced during inflation leave a distinctive imprint on the squeezed primordial bispectrum, characterized by power-law and oscillatory features, depending on the particle's mass and spin~\cite{Arkani-Hamed:2015bza,Assassi:2012zq,Chen:2009zp,Lee:2016vti}. Searching for these signatures forms the basis of the ``Cosmological Collider Physics" program~\cite{Arkani-Hamed:2015bza}, which aims to uncover the particle spectrum during inflation. In the past decade, significant theoretical progress has been made towards understanding and generalizing the Cosmological Collider scenario~\cite{Chen:2009zp, Chen:2009we, Baumann:2011nk, Chen:2012ge, Noumi:2012vr, Assassi:2012zq, Ghosh:2014kba, Arkani-Hamed:2015bza, Dimastrogiovanni:2015pla, Lee:2016vti, Kehagias:2017cym, An:2017hlx, Kumar:2017ecc, Baumann:2017jvh, Bordin:2018pca, Kumar:2018jxz, Goon:2018fyu, Hook:2019zxa, Kumar:2019ebj, Liu:2019fag, Wang:2019gbi, Bodas:2020yho, Lu:2021wxu, Pinol:2021aun, Cui:2021iie, Reece:2022soh, Chen:2022vzh, Chen:2023txq, Jazayeri:2023xcj, Chakraborty:2023qbp, Chakraborty:2023eoq, McCulloch:2024hiz, Quintin:2024boj}. Conversely, the observational imprints of the Cosmological Collider have only recently begun to be explored~\cite{Schmidt:2015xka, Gleyzes:2016tdh, MoradinezhadDizgah:2017szk, Cabass:2018roz, MoradinezhadDizgah:2018ssw, Akitsu:2020jvx, Green:2023uyz}. The first dedicated searches for the collider signal in cosmological data were conducted in the past year~\cite{Green:2023uyz, Cabass:2024wob, Sohn:2024xzd}. Thus, there is room for improvement on the observational front to optimally search for evidence of the Cosmological Collider using upcoming CMB and LSS surveys.

In this work, we present a numerical investigation of the imprint of intermediate-mass scalars\footnote{We use the term intermediate-mass to refer to particles in the complementary series, $0\leq \frac{m}{H}<\frac{3}{2}$, which imprint a power-law scaling in the squeezed bispectrum (see Eq.~\eqref{eq:squeezed_bk_qsfi}). In title-space, we prefer ``massive-ish particles."} on LSS correlation functions.  Specifically, we run $N$-body simulations with the collider squeezed bispectrum for two different particle masses. Our simulations use the same settings as the \textsc{Quijote-PNG} simulations~\cite{Coulton:2022rir}, but modified initial conditions. Using the separate universe formalism, we derive non-perturbative models for the squeezed matter bispectrum and collapsed matter trispectrum associated with this type of PNG, which are then validated deep into the non-linear regime using the simulations. We also use our simulations to study the large-scale halo bias for this type of PNG. These results help pave the way towards leveraging simulations and analytic models to search for signatures of the Cosmological Collider with LSS surveys.

The remainder of this paper is structured as follows. In Sec.~\ref{Sec:background}, we review the squeezed bispectrum in the Cosmological Collider scenario. We also derive theoretical models for the squeezed matter bispectrum, collapsed matter trispectrum, and scale-dependent bias associated with this scenario. In Sec.~\ref{Sec:methodology}, we discuss our simulations and analysis pipeline. We validate our models and simulations in Sec.~\ref{sec:results}. We conclude in Sec.~\ref{Sec:conclusions}. The appendices give useful theoretical results. In Appendix~\ref{sec:App_spin_and_osc}, we present an algorithm for generating $N$-body initial conditions with the Cosmological Collider bispectrum, including spin and oscillations. In Appendix~\ref{sec:App_estimators}, we derive the bispectrum and trispectrum estimators used in this work.

\begin{figure}[!t]
\centering
\includegraphics[width=0.99\linewidth]{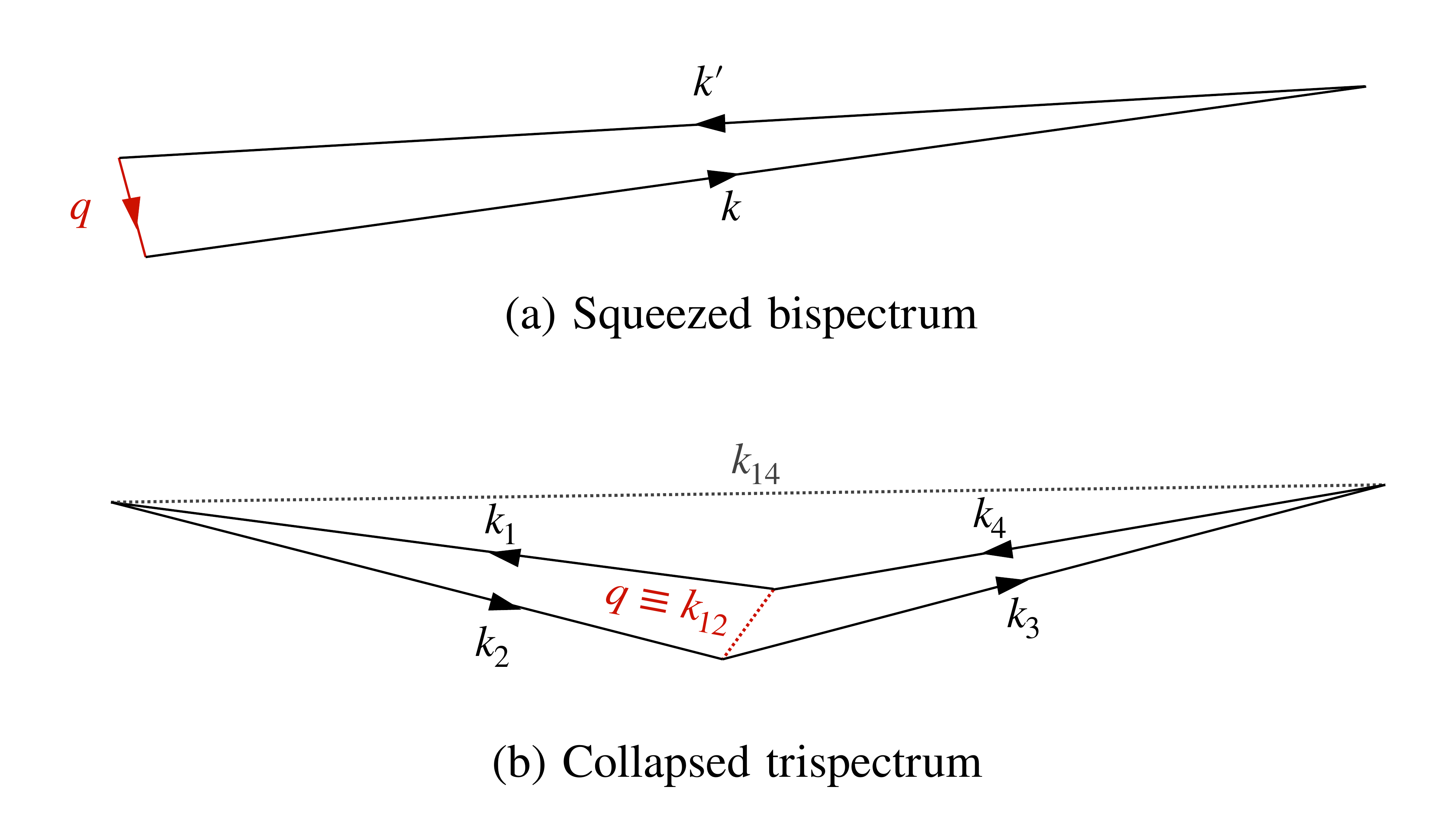}
\caption{The shape of the squeezed bispectrum (top) and collapsed trispectrum (bottom) in momentum space. The squeezed bispectrum is characterized by three external momenta, one of which, $q$, is much smaller than the other two, $k\approx k'$. The trispectrum is parameterized by four external momenta and two internal momenta. In the collapsed limit, one of the internal momenta, $q\equiv k_{12}$, is much smaller than all external momenta. Since the soft mode is \textit{internal}, the collapsed trispectrum is much less susceptible to cosmic variance.} \label{fig:squeezed_B_collapsed_T_shape_dependence}
\end{figure}

\section{Theoretical background}\label{Sec:background}
\subsection{Conventions}

\noindent We work in natural units, $c = 1$. We assume a flat $\Lambda$CDM fiducial cosmology based on the fiducial $\textsc{Quijote}$ simulations~\cite{Villaescusa-Navarro:2019bje}: $\Omega_m=0.3175$, $\Omega_{b,0}=0.049$, $h=0.6711$, $n_s=0.9624$, and $\sigma_8=0.834.$  We assume that all quantities are evaluated at a fixed redshift, $z$, which we will typically omit in the expressions below.

We use $\delta_m(\bx)\equiv\rho_m(\bx)/\bar{\rho}_m-1$ to denote the matter density fluctuation and $\delta_h(\bx)\equiv n_h(\bx)/\bar{n}_h-1$ to denote the halo number density fluctuation. We typically work with the Fourier-space fields, which are related to the configuration-space fields by
\begin{equation}
    \delta({\bx})= \int \frac{d^3k}{(2\pi)^3}\,\delta(\vk)e^{-i\vk \cdot \bx}\equiv \int_{\vk}\delta(\vk)e^{-i\vk \cdot \bx}.
\end{equation}
For a three-dimensional field $\delta(\vk)$,  the power spectrum, bispectrum, and trispectrum are defined by
\begin{align}
    \langle \delta(\vk_1)\delta(\vk_2)\rangle'_{c} &\equiv P(k_1)\,, \\
    \langle \delta(\vk_1) \delta(\vk_2) \delta(\vk_3) \rangle'_{c} &\equiv B(k_1, k_2,k_3)\,,\\
        \langle \delta(\vk_1) \delta(\vk_2) \delta(\vk_3)\delta(\vk_4) \rangle'_{c} &\equiv T(k_1, k_2,k_3,k_4, k_{12},k_{14})\,,
\end{align}
respectively, where $\langle \dots \rangle'_c$ denotes the connected correlator with the overall momentum-conserving delta function stripped off. The bispectrum is parameterized by the side lengths of three momentum vectors forming a triangle. Similarly, the trispectrum is parameterized by the lengths of the four sides and two diagonals spanning a tetrahedron. Finally, we define the sum $\vk_{12}\equiv \vk_1+\vk_2.$

In this work, we are interested in soft limits of correlation functions, wherein one of the momenta is significantly smaller than all others. We will always use $q$ to denote the soft mode. For the bispectrum, we analyze the ``squeezed limit," where one of the external momenta, $q,$ is much smaller than the other two, $k\approx k'$. For convenience, we will often write the bispectrum in terms of two of the three momenta as $B(\vk_1,\vk_2)$. For the trispectrum, we focus on the ``collapsed limit," where one of the internal momenta, $q\equiv k_{12}$, is much smaller than all external momenta. See Fig.~\ref{fig:squeezed_B_collapsed_T_shape_dependence} for a graphical representation of these limits.

\subsection{Cosmological Collider bispectrum}
\noindent The main goal of this work is to study LSS correlators in cosmologies with the following squeezed primordial bispectrum,
\begin{equation}\label{eq:squeezed_bk_qsfi}
\lim_{q\ll k}B_{\Phi}(\vq,\vk) = 4 f_{\rm NL}^{\Delta}\left(\frac{q}{k}\right)^\Delta P_{\Phi}(q)P_{\Phi}(k),
\end{equation}
where $f_{\rm NL}^\Delta$ is the amplitude of PNG,\footnote{We normalize $f_{\rm NL}^\Delta$ such that $f_{\rm NL}^{\Delta=0}=f_{\rm NL}^{\rm loc.}$. Note that this differs by a factor of three from the convention used in~\citet{Green:2023uyz}, where their $f_{\rm NL}^{\Delta}$ is defined such that $f_{\rm NL}^{\Delta=2}=f_{\rm NL}^{\rm eq.}$, thus $f_{\rm NL}^{\rm loc.}=3f_{\rm NL}^{\Delta=0}$.} $\Delta$ is a (possibly complex) exponent, and $P_{\Phi}$ is the power spectrum of the primordial Bardeen potential, $\Phi$. Here, $q$ is the ``soft" mode and $k\approx k'$ are the ``hard" modes (see Fig.~\ref{fig:squeezed_B_collapsed_T_shape_dependence}).  

The squeezed bispectrum in Eq.~\eqref{eq:squeezed_bk_qsfi} generically arises if there are scalars\footnote{The discussion in this section can be extended to spin-$s$ particles. In this case, the squeezed bispectrum takes the following anisotropic form~\cite{Arkani-Hamed:2015bza},
\begin{equation}\label{eq:squeezed_bk_qsfi_spin}
\lim_{q\ll k}B_{\Phi}(\vq,\vk) = 4 f_{\rm NL}^{\Delta}\left(\frac{q}{k}\right)^\Delta \mathcal{L}_s(\hat{\vq}\cdot\hat{\vk})P_{\Phi}(q)P_{\Phi}(k),
\end{equation}
where $\mathcal{L}_s(x)$ is the Legendre polynomial of order $s.$ The exponent depends on both the spin and mass as follows,
\begin{equation}\label{eq:exponent_spin}
    \Delta=3/2-\sqrt{\left(s-{1}/{2}\right)^2-m^2/H^2} \textrm{ for }s\neq 0.
\end{equation}
We leave a study of spinning particles to a future work.} with mass comparable to the Hubble scale during inflation~\cite{Arkani-Hamed:2015bza}. In this case, the exponent is set by the mass of the field,
\begin{equation}\label{eq:exponent_scalar}
    \Delta=3/2-\sqrt{9/4-m^2/H^2},
\end{equation}
which has three interesting regimes:
\begin{enumerate}
    \item \textbf{Light fields}:  For light fields ($m\ll H$), $\Delta\approx 0$, and we recover the \emph{local model} associated with multi-field inflation models parameterized by $f_{\rm NL}^{\rm loc.}$~\cite{Komatsu:2001rj}.
    \item \textbf{Intermediate-mass fields}: For intermediate masses ($0\ll m\leq 3H/2$; complementary series), $0<\Delta\leq 3/2$, and the squeezed bispectrum includes a characteristic \emph{power-law} scaling. This squeezed bispectrum has been analyzed in several specific models of inflation, most notably quasi-single-field inflation (QSFI)~\cite{Chen:2009we, Chen:2009zp, Baumann:2011nk, Assassi:2012zq, Noumi:2012vr, Sefusatti:2012ye}; however, it is a generic consequence of the presence of intermediate-mass particles during inflation~\cite{Arkani-Hamed:2015bza}.
    \item \textbf{Massive fields}: For heavier fields ($m>3H/2$; principal series), $\Delta$ is complex and there are \emph{oscillations} in the squeezed bispectrum. These oscillations produce interesting signatures in LSS correlators~\cite{Cyr-Racine:2011bjz, MoradinezhadDizgah:2017szk,  MoradinezhadDizgah:2018ssw, Cabass:2018roz}. 
\end{enumerate}

In what follows, we derive theoretical predictions for the squeezed matter bispectrum, collapsed matter trispectrum, and scale-dependent bias sourced by the primordial bispectrum in Eq.~\eqref{eq:squeezed_bk_qsfi} for real values of $\Delta$. We defer an investigation of oscillatory bispectra to a future work. We work in the peak-background split formalism and follow a procedure similar to that used previously to derive the scale-dependent bias effect for non-local PNG~\cite{Schmidt:2010gw, Scoccimarro:2011pz, Schmidt:2012ys, Desjacques:2016bnm,  Cabass:2018roz}. 

\subsection{Non-perturbative model for the squeezed matter bispectrum}

\noindent First, we derive a non-perturbative model for the squeezed matter bispectrum. Our model generalizes the model derived for $f_{\rm NL}^{\rm loc.}$ in Ref.~\cite{Goldstein:2022hgr} to account for the $(q/k)^\Delta$ momentum dependence in Eq.~\eqref{eq:squeezed_bk_qsfi}. The central idea is to split the (squeezed) bispectrum into a term corresponding to the gravitational-evolved contribution sourced by PNG and a term corresponding to the contributions from gravitational non-Gaussianity that would be present even in the absence of $f_{\rm NL}^\Delta,$ \emph{i.e.},
\begin{equation}
    B_{m}(\vq,\vk)=B_{m}^{ f_{\rm NL}^\Delta}(\vq,\vk)+B_{m}^{\rm grav.}(\vq,\vk).
\end{equation}
This is possible solely because the leading-order contribution to the bispectrum associated with the squeezed primordial bispectrum in Eq.~\eqref{eq:squeezed_bk_qsfi} is proportional to $P_m(q)/q^{2-\Delta}$, which \emph{cannot be generated by gravitational evolution} because of the LSS consistency relation, Eq.~\eqref{eq:LSS_CR_statement}.

\subsubsection{Primordial non-Gaussianity contribution}\label{subsec:squeezed_b_primordial_SU}

\noindent In the squeezed limit, the matter bispectrum can be described by the correlator $\av{\delta_{ m,L}(\vq) P_{m}(k|\bx)|_{\Phi_L}}'$ where $\delta_{ m, L}(\bm q)$ is the long-wavelength (linear) matter density field and $P_{m}(k|\bx)|_{\Phi_L}$ is the locally measured small-scale (non-linear) matter power spectrum at position $\bx$ in the presence of a long-wavelength potential $\Phi_L$. For $q \ll k$, we can treat the long mode as a background and expand the locally measured small-scale power spectrum as 
\begin{equation}\label{eq:Pk_expansion}
        P_m(k|\bx)|_{\Phi_L} ={} P_m(k)+\int_{\vp} \, \frac{\partial P_m(k)}{\partial\,\Phi_{L}(\vp)}\, \Phi_{L}(\vp) + \dots \,
\end{equation}
This yields the squeezed matter bispectrum
\begin{align}\label{eq: squeezed_bispectrum_integral}
    \begin{split}
        B_m(\vq,\vk) &= \int_{\vp} \frac{\partial P_m(k)}{\partial\,\Phi_{L}(\vp )}\av{\delta_{m,L}(\vq) \Phi_{L}(\vp)}\,, \\
        &=\frac{\partial P_m(\vk)}{\partial\,\Phi_{L}(\vq)}\frac{P_m(q)}{\mathcal{M}(q,z)} \,.
    \end{split}
\end{align}
Here, we used Poisson's equation to relate the linear density field to the long-wavelength gravitational potential,
\begin{equation*}
    \delta_{m,L}(\vq, z)=\frac{2\,q^2\,T(q)D_{\rm md}(z)}{3\Omega_mH_0^2}\Phi_L(\vq)\equiv \mathcal{M}(q,z)\Phi_L(\vq),
\end{equation*}
where $T(q)$ is the transfer function normalized to unity on large scales and $D_{\rm md}(z)$ is the linear growth factor normalized to the scale factor during matter domination, $\Omega_m$ is the present-day matter density, and $H_0$ is the Hubble constant.

To proceed, we need to evaluate the potential derivative in Eq.~\eqref{eq: squeezed_bispectrum_integral}. Since the derivative acts on the \emph{non-linear} matter power spectrum, it is challenging to compute using conventional techniques, such as perturbation theory. Instead, the derivative can be estimated using the separate universe approach~\cite{Wagner:2014aka}, wherein we express the derivative in terms of a change in the background cosmology to mimic the primordial mode coupling described by Eq.~\eqref{eq:squeezed_bk_qsfi}. Specifically, according to Eq.~\eqref{eq:squeezed_bk_qsfi} a long-wavelength potential fluctuation on scale $\vq$ induces a scale-dependent modification of the small-scale linear density field~\cite{Desjacques:2016bnm, Cabass:2018roz},
\begin{equation}\label{eq:delta_modification_SU}
   \delta_{m,L}(\boldsymbol{k}|\Phi_L(\boldsymbol{q}))=\left(1+2f_{\rm NL}^\Delta\left(\frac{q}{k}\right)^\Delta\Phi_L(\boldsymbol{q})  \right)\delta_{m,L}(\boldsymbol{k}).
\end{equation}

\begin{figure}[!t]
\centering
\includegraphics[width=0.99\linewidth]{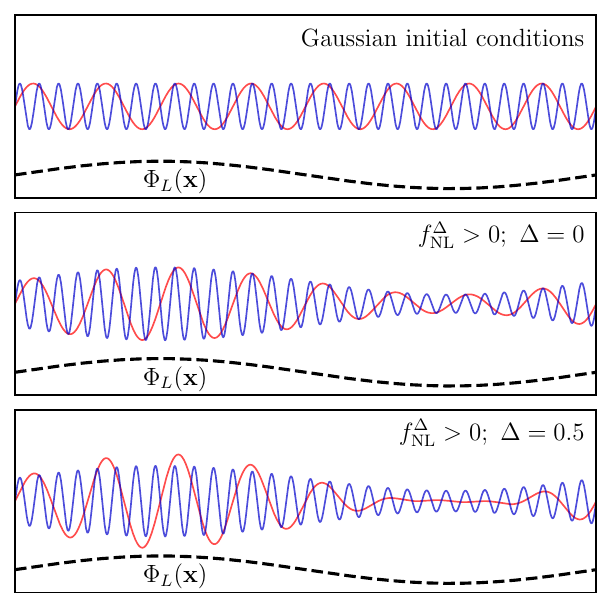}
\caption{Graphical representation of the modulation of two small-scale density fluctuations with \emph{fixed amplitude, but different momenta,} to a long-wavelength gravitational potential, $\Phi_L(\bx)$ (black-dashed) as described by  Eq.~\eqref{eq:delta_modification_SU}. For Gaussian initial conditions (top), the small-scale modes are uncorrelated with $\Phi_L(\bx)$. For local non-Gaussianity (middle), the small-scale fluctuations respond uniformly to a long-wavelength potential. This is equivalent to changing $\sigma_8$. Finally, for the intermediate-mass particle scenario (bottom), the modulation depends on the momenta of the small-scale modes. } \label{fig:mode_response}
\end{figure}

Before continuing, we provide a graphical depiction of this mode coupling for several values of $f_{\rm NL}^\Delta$ and $\Delta$. In Fig.~\ref{fig:mode_response}, we show the response of two small-scale density fluctuations with fixed amplitude, but different momenta, to a long-wavelength potential fluctuation for several values of $f_{\rm NL}^\Delta$ and $\Delta$. For Gaussian initial conditions, the small-scale modes are uncorrelated with the long-wavelength potential. For local non-Gaussianity, $\Delta=0$, the small-scale response is uniform. In this case, the modulation is equivalent to changing the amplitude of clustering, $\sigma_8.$ Finally, if $\Delta\neq 0$, then the modulation of the small-scale fluctuations depends on their momenta.

Per Eq.~\eqref{eq:delta_modification_SU}, the locally measured, small-scale linear matter power spectrum is modulated by the long-wavelength potential, $\Phi_L(\vq)$, as follows,
\begin{equation}\label{eq:local_power_response}
    P_{m}^{\rm lin.}(k|\bx)|_{\Phi_L}=\left[1+4f_{\rm NL}^{\Delta}\left(\frac{q}{k}\right)^{\Delta}\Phi_L(\vq)e^{i\vq\cdot\bx}\right]P_{m}^{\rm lin.}(k).
\end{equation}
Treating $\vq$ as a background, the mode coupling in Eq.~\eqref{eq:squeezed_bk_qsfi} is equivalent to a scale-dependent modification of the linear power spectrum,
\begin{equation}\label{eq:Pk_lin_modification_SU}
    P_{m}^{\rm lin.}(k|\epsilon,\Delta)\equiv \left(1+2\epsilon k^{-\Delta} \right)P_{m}^{\rm lin.}(k)
\end{equation}
for small parameter $\epsilon$ encoding the long-wavelength field. Thus, the potential derivative is
\begin{equation}\label{eq:potential_derivative} 
     \frac{\partial P_m(k)}{\partial\,\Phi_{L}(\vq  )}=2f_{\rm NL}^\Delta q^\Delta \frac{\partial P_m(k|\epsilon, \Delta)}{\partial \epsilon }\bigg\vert_{\epsilon=0},
\end{equation}
where $\partial P_m(k|\epsilon, \Delta)/\partial \epsilon \vert_{\epsilon=0}$ is the derivative of the non-linear matter power spectrum with respect to a change in the initial linear power spectrum described by Eq.~\eqref{eq:Pk_lin_modification_SU}. Since PNG corresponds to mode coupling in the initial, \emph{i.e.}, linear density field, Eq.~\eqref{eq:potential_derivative} can be evaluated in the non-linear regime using simulations with modified initial conditions. We shall elaborate on this point later.

 Using Eq.~\eqref{eq:potential_derivative}, a non-perturbative model for the primordial contribution to the squeezed matter bispectrum at leading order in $f_{\rm NL}^\Delta$ is
 \begin{align}\label{eq:squeezed_B_non_pert_collider_model}
     B^{f_{\rm NL}^\Delta}_m(\vq, \vk) &= \frac{2f_{\rm NL}^{\Delta}q^\Delta }{\mathcal{M}(q,z)}P_{m}(q)\frac{\partial P_m(k|\epsilon, \Delta)}{\partial \epsilon}\bigg\vert_{\epsilon=0},\\
 &= \frac{3f_{\rm NL}^\Delta\Omega_mH_0^2}{D_{\rm md}(z)}\frac{P_m(q)}{q^{2-\Delta}T(q)}\frac{\partial P_m(k|\epsilon, \Delta)}{\partial \epsilon}\bigg\vert_{\epsilon=0}. \nonumber
\end{align}

We now make a couple of comments about Eq.~\eqref{eq:squeezed_B_non_pert_collider_model}. First, on large scales, the ratio $B^{f_{\rm NL}^\Delta}_m(\vq,\vk)/P_m(q)\propto 1/q^{2-\Delta}$ (with $0\leq\Delta\leq 3/2$). This term cannot be generated by gravitational evolution alone because it violates the LSS consistency relation, Eq.~\eqref{eq:LSS_CR_statement}. As a result, we can model the squeezed bispectrum contributions sourced by gravitational non-Gaussianity separately from those sourced by the Cosmological Collider.\footnote{This also applies for the bispectrum templates associated with particles mass $m>3H/2$, as well as particles with spin.} Second, by expressing the potential derivative in terms of $\epsilon,$ we have written the bispectrum in a form that is of practical use. In particular, we can compute the derivative in Eq.~\eqref{eq:squeezed_B_non_pert_collider_model} by running simulations with a modified initial linear power spectrum according to Eq.~\eqref{eq:Pk_lin_modification_SU} for small values of $\pm \epsilon$ and computing $\partial P_m(k|\epsilon,\Delta)/\partial \epsilon|_{\epsilon=0}$ numerically from the simulation output. We describe this procedure in more detail in Sec.~\ref{subsec:separate_universe}.

Finally, we check that Eq.~\eqref{eq:squeezed_B_non_pert_collider_model} produces the expected behavior in known limits. At tree level, we can compute the potential derivative from Eq.~\eqref{eq:Pk_lin_modification_SU}, leading to
\begin{equation}
    B_m^{\rm tree}(\vq, \vk)=\frac{4 f_{\rm NL}^\Delta }{\mathcal{M}(q,z)}\left(\frac{q}{k} \right)^\Delta P_m(q)P_m(k),
\end{equation}
in agreement with Eq.~\eqref{eq:squeezed_bk_qsfi}. For local non-Gaussianity ($\Delta=0$) the transformation in Eq.~\eqref{eq:Pk_lin_modification_SU} is equivalent to rescaling $\sigma_8 \rightarrow \sigma_8(1+\epsilon)$. The potential derivative is then $\partial P_m(k)/\partial \epsilon \vert_{\epsilon=0}= 2\,\partial P_m(k)/\partial \log \sigma_8^2$ and we recover the non-perturbative model for the squeezed matter bispectrum in local PNG introduced in Ref.~\cite{Goldstein:2022hgr}.

\subsubsection{Gravitational contribution}

\noindent Even in the absence of PNG, non-linear structure formation generates a bispectrum. On sufficiently large scales, any primordial contribution of the form given by Eq.~\eqref{eq:squeezed_B_non_pert_collider_model} will dominate the equal-time squeezed bispectrum by the LSS consistency relations. Nevertheless, for the range of scales probed by cosmological surveys (and simulations) and for the values of $f_{\rm NL}^\Delta\lesssim\mathcal{O}(100)$~\cite{Sohn:2024xzd} that are allowed with current observations, the gravitational contribution is typically non-negligible. Thus, to obtain unbiased constraints on $f_{\rm NL}^\Delta$ from the squeezed matter bispectrum, we need a model also for the signal in the absence of PNG.

Whereas the full matter bispectrum is difficult to model analytically beyond perturbative scales, the squeezed matter bispectrum can be modeled non-perturbatively using the response function approach~\cite{Valageas:2016hhr, Wagner:2014aka, Chiang:2014oga, Wagner:2015gva, Chiang:2017vsq, Esposito:2019jkb, Barreira:2017sqa, Biagetti:2022ckz}, \emph{i.e.}, 
\begin{align}\label{eq:b_3D_grav}
     B_m^{\rm grav.}(\bm q,\bm k)=&\,\left[a_0(k)+{a}_2(k)\frac{q^2}{k^2}+\cdots\right]P(k)P(q)\,,
\end{align}
where $a_0$ and $a_2$ are coefficients characterizing the response of small-scale matter clustering to a long-wavelength density perturbation.  In principle, these coefficients can be estimated from simulations (\emph{e.g.}, ~\cite{Chiang:2017jnm, Biagetti:2022ckz}) or estimated from theory (\emph{e.g.},~\cite{Valageas:2013zda, Nishimichi:2014jna}); however, we follow the approach in Refs.~\cite{Esposito:2019jkb, Goldstein:2022hgr}, where we treat the coefficients as free parameters that we marginalize over.\footnote{This definition of $a_0$ and $a_2$ differs slightly from the definition used in Ref.~\cite{Goldstein:2022hgr}, where we absorbed any factors of $k$ and $P(k)$ into the coefficients themselves. The two approaches are equivalent, but the convention adopted here has dimensionless coefficients.} In practice, we measure the squeezed bispectrum averaged over modes with $k_{\rm min}<k,k'<k_{\rm max}$; hence the response functions reduce to scalars $\bar{a}_0$ and $\bar{a}_2$. We describe this procedure in more detail in Sec.~\ref{subsec:measurements_modelling_likelihood}.

Combining the contributions due to PNG, Eq.~\eqref{eq:squeezed_B_non_pert_collider_model}, and non-linear structure formation, Eq.~\eqref{eq:b_3D_grav}, our model for the squeezed matter bispectrum is 
\begin{widetext}
\begin{equation}\label{eq:squeezed_bispectrum_non_pert}
   B_m(q,k)= \frac{3f_{\rm NL}^\Delta\Omega_mH_0^2}{D_{\rm md}(z)}\frac{P_m(q)}{q^{2-\Delta}T(q)}\frac{\partial P_m(k|\epsilon, \Delta)}{\partial \epsilon}\bigg\vert_{\epsilon=0}+\left[a_0(k)+a_2(k)\frac{q^2}{k^2}\right]P(q)P(k)+\mathcal{O}\left((f_{\rm NL}^\Delta)^2,q^4/k^4\right).
\end{equation}
\end{widetext}

\subsection{Non-perturbative model for the collapsed matter trispectrum}
\noindent Having derived a non-perturbative model for the squeezed matter bispectrum, we now turn to the collapsed matter trispectrum. In this limit, the trispectrum is a function of the magnitude of the soft mode, $q\equiv k_{12}$, as well as two external momenta $k_1\approx k_2$ and $k_3\approx k_4$ (see Fig.~\ref{fig:squeezed_B_collapsed_T_shape_dependence}). In the separate universe picture, the collapsed trispectrum contains the following correlator $\av{ P_{m}(k_1|\bx)|_{\Phi_L} P_{m}(k_3|\bx)|_{\Phi_L}}'$  (see, \emph{e.g.}, Refs.~\cite{Lewis:2011au, Kenton:2016abp}), which leads to the relation:
\begin{align}\label{eq:collapsed_T_no_contact}
    T_m(q,k_1,k_3)&=\frac{\partial P_m(k_1)}{\partial \Phi_L(\vq)}\frac{\partial P_m(k_3)}{\partial \Phi_L(\vq)}P_{\Phi}(q)\nonumber,\\
    &=\frac{B_m(q,k_1)B_m(q,k_3)}{P_m(q)},
\end{align}
where the second equality follows from Eq.~\eqref{eq: squeezed_bispectrum_integral}. Thus, given a non-perturbative model for the squeezed bispectrum, \emph{i.e.}, Eq.~\eqref{eq:squeezed_bispectrum_non_pert}, one can readily derive a non-perturbative model for the collapsed trispectrum (though see the discussion below). In practice, gravitational evolution or alternative inflationary scenarios, such as $g_{\rm NL}$ PNG, produce contributions to the trispectrum that are independent of the internal mode, $q$. For non-zero $f_{\rm NL}^\Delta$, these contributions are generically suppressed on the largest scales, where the collapsed trispectrum is dominated by the $(f_{\rm NL}^{\Delta})^{2} P_m(q)/q^{4-2\Delta}$ contribution. Nevertheless, the contact terms are non-negligible for the range of scales analyzed in this work. To account for these contributions, we model the collapsed trispectrum as,\footnote{For the models considered here, this equality holds on a realization-by-realization basis. Thus, neglecting contact terms, the collapsed $T$ is set by the \emph{realization} of the squeezed $B$ and soft $P$.}
\begin{equation}\label{eq:trispectrum_no_tauNL}
   T_m(q,k_1,k_3)= \frac{B_m(q,k_1)B_m(q,k_3)}{P_m(q)}+T_0(k_1,k_3),
\end{equation}
where $B_m(q,k)$ is the squeezed bispectrum, given by Eq.~\eqref{eq:squeezed_bispectrum_non_pert}, and $T_0(k_1,k_3)$ encapsulates all $q$-independent contributions to the collapsed trispectrum. Finally, we measure the trispectrum integrated over a wide range of hard modes; thus $T_0(k_1,k_3)$ reduces to a scalar, $\bar{T}_0$, which we marginalize over.

In some inflation models, the collapsed trispectrum can receive contributions from exchange diagrams beyond the squeezed bispectrum. In the local model, these contributions are conventionally parameterized by the amplitude $\tau_{\rm NL}^{\rm loc.}$.\footnote{ We focus on $\tau_{\rm NL}^{\rm loc.}$ for simplicity, but these results can be easily extended to non-zero values of $\Delta$ by defining an amplitude $\tau_{\rm NL}^\Delta$. In full, such additional contributions have a rich phenomenology, tracing, for example, scalars that do not couple quadratically to the inflaton and non-longitudinal spin-states. We ignore such effects in this work.} A non-perturbative model for the collapsed trispectrum including $f_{\rm NL}^{\rm loc.}$ and $\tau^{\rm loc.}_{\rm NL}$ is
\begin{widetext}
\begin{equation}\label{eq:tau_NL_collapsed_T}
     T^{\rm loc.}_m(q,k_1,k_3)= \left(\frac{25}{36}\tau^{\rm loc.}_{\rm NL}-(f_{\rm NL}^{\rm loc.})^2\right)\frac{36\,\Omega_m H_0^2 }{D_{\rm md}(z)^2}\frac{P_{m}(q)}{q^4T^2(q)}\frac{\partial P_m(k_1)}{\partial \log\sigma_8^2}\frac{\partial P_m(k_3)}{\partial \log \sigma_8^2}
     +\frac{B_m(q,k_1)B_m(q,k_3)}{P_m(q)}+T_0(k_1,k_3),
\end{equation}
\end{widetext}
where the factor of $6/5$ is conventional and $B_m(q,k)$ is computed from Eq.~\eqref{eq:squeezed_bispectrum_non_pert} with $\Delta=0.$ 

For local non-Gaussianity described by a quadratic rescaling of the primordial potential ($\Phi\to \Phi+f_{\rm NL}^{\rm loc.}\Phi^2$), the collapsed trispectrum is completely determined by the squeezed bispectrum, \emph{i.e.}, $\tau_{\rm NL}^{\rm loc.}=\left(\frac{6}{5}f_{\rm NL}^{\rm loc.}\right)^2$. In contrast, for inflationary scenarios in which multiple fields contribute to the curvature perturbation, the amplitude of the collapsed trispectrum will be boosted. Regardless of the inflationary model, $\tau_{\rm NL}^{\rm loc.}$ and $f_{\rm NL}^{\rm loc.}$ must satisfy the Suyama-Yamaguchi inequality (\emph{e.g.},~\cite{Suyama:2007bg, Sugiyama:2011jt, 2011PhRvL.107s1301S, Baumann:2012bc, Assassi:2012zq}),
\begin{equation}
    \tau_{\rm NL}^{\rm loc.}\geq \left( \frac{6}{5}f_{\rm NL}^{\rm loc.}\right)^2.
\end{equation}
\subsection{Scale-dependent bias}

\noindent In addition to modifying higher-order correlation functions of the matter field, the squeezed primordial bispectrum in Eq.~\eqref{eq:squeezed_bk_qsfi} impacts the two-point statistics of biased tracers, such as galaxies and halos. Specifically, the correlation between the long-wavelength primordial potential and the locally measured small-scale power spectrum induces spatial variations in the halo number density, $n_h(\bx)$. These variations are correlated with the long-wavelength potential. Thus, the large-scale clustering of biased tracers is a powerful probe of inflationary scenarios with a large squeezed primordial bispectrum. A prime example of this is the ``scale-dependent bias" in local PNG~\cite{Dalal:2007cu, Matarrese:2008nc, Slosar:2008hx, Desjacques:2008vf}, where the large-scale bias includes a term proportional to $f_{\rm NL}^{\rm loc.}/q^2.$ More general types of PNG, including the Cosmological Collider bispectrum, induce their own scale-dependent bias~\cite{Schmidt:2010gw, Scoccimarro:2011pz, PhysRevD.84.061301, PhysRevD.84.063512,  Schmidt:2012ys, Desjacques:2016bnm,  Gleyzes:2016tdh, Cabass:2018roz, Green:2023uyz}. Here, we review the scale-dependent bias associated with the squeezed bispectrum in Eq.~\eqref{eq:squeezed_bk_qsfi}. For a more detailed treatment, we refer the reader to Refs.~\cite{Schmidt:2010gw, PhysRevD.84.061301, PhysRevD.84.063512, Schmidt:2012ys, Desjacques:2016bnm, Cabass:2018roz}.

On large scales, the halo auto-spectrum and halo-matter cross-spectrum are given by
\begin{equation}\label{eq:Phh}
P_{ h}(q)=\left[ b_1+\frac{b_{\Psi,\Delta} f_{\rm NL}^{\Delta}q^\Delta}{\mathcal{M}(q)} \right]^2P_{ m}(q)+\frac{1}{\bar{n}_h}(1+\epsilon_N),
\end{equation}
\begin{equation}\label{eq:Phm}
P_{ hm}(q)=\left[ b_1+\frac{b_{\Psi,\Delta} f_{\rm NL}^{\Delta}q^\Delta}{\mathcal{M}(q)} \right]P_{m}(q),
\end{equation}
respectively, where $b_1$ is the linear halo bias, $b_{\Psi,\Delta}$ is a non-Gaussian bias parameter, $\bar{n}_h$ is the mean number density of halos, and $\epsilon_N$ is a small parameter that encapsulates the leading order effect of non-Poissonian shot noise. On large scales, the bias is proportional to $f_{\rm NL}^\Delta/q^{2-\Delta}$ (and physically encodes the same mechanism as the above squeezed bispectrum). Thus, in principle, one can constrain $f_{\rm NL}^\Delta$ using the large-scale halo power spectrum. In practice, since $f_{\rm NL}^\Delta$ is completely degenerate with the non-Gaussian bias parameter, $b_{\Psi,\Delta}$, an estimate of $b_{\Psi,\Delta}$ is necessary to robustly constrain $f_{\rm NL}^\Delta$ using the scale-dependent bias~\cite{Reid_2010, Barreira:2020kvh, Barreira:2021ueb, Lazeyras:2022koc,Barreira:2022sey, Barreira:2023rxn}. In the remainder of this section, we discuss several ways to estimate $b_{\Psi,\Delta}$ that we will test with $N$-body simulations in Sec.~\ref{subsec:res_scale_dep_bias}.

\subsubsection{Separate universe bias parameters}
\noindent In the peak-background split formalism, the non-Gaussian bias parameter $b_{\Psi,\Delta}$ describes the response of galaxy formation to a long-wavelength potential perturbation. Specifically, for the squeezed bispectrum in Eq.~\eqref{eq:squeezed_bk_qsfi}, the non-Gaussian bias is (\emph{e.g.},~\cite{Desjacques:2016bnm})
\begin{equation}\label{eq:b_psi_SU}
    b_{\Psi,\Delta} = \frac{2\,\partial \log  \bar{n}_h}{\partial \epsilon}\bigg\vert_{\epsilon=0},
\end{equation}
where $\bar{n}_h$ is the mean number density of halos and the derivative is taken with respect to a change in the linear power spectrum according to Eq.~\eqref{eq:Pk_lin_modification_SU}. Physically, Eq.~\eqref{eq:b_psi_SU} can be understood as the analog of the separate universe potential derivative (Eq.~\ref{eq:potential_derivative}) that we derived for the squeezed matter bispectrum model. Indeed, $b_{\Psi,\Delta}$ can be estimated using the same set of separate universe simulations used to compute Eq.~\eqref{eq:potential_derivative}. For local PNG, $\Delta=0$, Eq.~\eqref{eq:b_psi_SU} becomes $b_{\Psi,\Delta}=4\,\partial \log \bar{n}_h/\partial \log \sigma_8^2$, which is consistent with the separate universe simulations used in previous studies of local PNG, \emph{e.g.},~\cite{Barreira:2020kvh, Hadzhiyska:2024kmt}.

\subsubsection{Universal mass functions}

\noindent Assuming a universal halo mass function (UMF),\footnote{In this context, universal means that the halo mass function depends only on the mean density of the Universe, the variance of the density field smoothed on the scale of the Lagrangian radius of the halo, $R_*(M)$, and the Jacobian of the transformation relating the smoothed variance to the halo mass, $J\equiv |\partial \log \sigma_{R_*}/\partial M|.$ See Sec. 7.3 of Ref.~\cite{Desjacques:2016bnm} for more details.} the non-Gaussian bias is (\emph{e.g.},~\cite{Desjacques:2016bnm})
\begin{equation}\label{eq:non_Gaussian_bias_UMF}
    b_{\Psi,\Delta}^{\rm UMF}=\left[b_\phi^{\rm UMF}+4\left(\frac{d\log \sigma_{R_*,-\Delta/2}^2}{d\log \sigma_{R_*}^2}-1\right) \right]\frac{\sigma_{R_*,-\Delta/2}^2}{\sigma_{R_*}^2},
\end{equation}
where $b_\phi^{\rm UMF}=2\delta_c(b_1-1)$ is the universality prediction for the local PNG bias parameter and $\delta_c=1.686$ is the threshold for spherical collapse linearly extrapolated to $z=0$. In Eq.~\eqref{eq:non_Gaussian_bias_UMF}, $R_*\equiv R_*(M)$ denotes the Lagrangian radius of the halo, and $\sigma_{R,p}$ is the momentum-weighted variance of the linear density field smoothed on some scale $R$. Specifically,
\begin{equation}\label{eq:spectral_moment}
    \sigma_{R,p}^2\equiv \int \frac{d^3k}{(2\pi)^3}k^{2p}W_R^2(k)P_{m}^{\rm lin.}(k),
\end{equation}
where $W_R(k)=3j_1(kR)/(kR)$ is the Fourier transform of a spherical top-hat filter with radius $R.$ 
 
Eq.~\eqref{eq:non_Gaussian_bias_UMF} is practically useful as it provides a means to compute $b_{\Psi,\Delta}$ directly from $b_1$ and the halo mass (which sets $R_*$)\footnote{The logarithmic derivative in Eq.~\eqref{eq:non_Gaussian_bias_UMF} can be computed numerically using finite differences with Eq.~\eqref{eq:spectral_moment} at $R_*\rightarrow R_*\pm \epsilon$.}. However, recent studies in the context of local PNG have shown that Eq.~\eqref{eq:non_Gaussian_bias_UMF} can be a poor approximation for realistic halo and galaxy samples~\cite{Barreira:2020kvh, Barreira:2021ueb, Lazeyras:2022koc,Barreira:2022sey, Barreira:2023rxn}. Thus, further work is required to understand when, if it all, Eq.~\eqref{eq:non_Gaussian_bias_UMF} can be applied to realistic galaxy surveys.

Finally, from Eq.~\eqref{eq:Phh}, $b_{\Psi,\Delta}$ has units of $({\rm Mpc}/h)^\Delta$. Thus, by dimensional analysis, the non-Gaussian bias parameter can be approximated as
\begin{equation}\label{eq:b_psi_dim_analysis}
    b_{\Psi,\Delta} \approx 2\delta_c(b_1-1)R_*^\Delta.
\end{equation}
This approximation for the non-Gaussian bias has been used in several recent works~\cite{Gleyzes:2016tdh, Green:2023uyz}.

\section{Methodology}\label{Sec:methodology}
\subsection{Simulations}

\noindent This section describes the details of the simulations used in this work. We run two types of simulations to analyze non-linear LSS in the Cosmological Collider scenario. First, we run simulations with PNG in the initial conditions described by the squeezed bispectrum in Eq.~\eqref{eq:squeezed_bk_qsfi} for several values of $\Delta$ and $f_{\rm NL}^\Delta$. Second, for each value of $\Delta$, we run separate universe simulations where we modify the input linear power spectrum according to Eq.~\eqref{eq:Pk_lin_modification_SU} for small values $\pm\epsilon$ to compute the non-linear responses in the bispectrum, trispectrum, and scale-dependent bias models. Table~\ref{tab:sim_parameters} summarizes the simulation parameters used in this work.

Aside from modifying the initial conditions, which we describe in the following subsections, we generate all simulations using the same settings as the \textsc{Quijote} suite~\cite{Villaescusa-Navarro:2019bje, Coulton:2022rir}.\footnote{\href{https://quijote-simulations.readthedocs.io/en/latest/}{https://quijote-simulations.readthedocs.io/en/latest/}} We refer the reader to Ref.~\cite{Villaescusa-Navarro:2019bje} for a detailed discussion of the \textsc{Quijote} simulations, but we review the main information here. All simulations are run using the \textsc{Gadget-3} treePM code~\cite{Springel:2005mi} and contain $512^3$ particles in a $(1~{\rm Gpc}/h)^3$ comoving volume. We assume a fiducial cosmology with $\Omega_m=0.3175$, $\Omega_{b}=0.049$, $h=0.6711$, $n_s=0.9624$, and $\sigma_8=0.834$ and compute all transfer functions using CAMB~\cite{Lewis:1999bs}. We produce halo catalogs using the Friends-of-Friends (FoF) algorithm~\cite{Davis:1985rj} with a linking length of $b=0.2.$

\begin{table}[!t]
    \centering

    \begin{tabular}{ |c|c|c| } 
    
     \multicolumn{3}{c}{~~\it{PNG Simulations}~~}          \\
     \hline
     \hline
   $~~f_{\rm NL}^\Delta~~$ & $~~\Delta~~$ & $N_{\rm sim}$ \\ 
    \hline
    100 &  0 &  50  \\
    \hline
    300 &  0.5  &  50  \\
    \hline
     1000 &  1  &  50  \\
     \hline
    \end{tabular}
    \quad
    \begin{tabular}{ |c|c|c| } 
         \multicolumn{3}{c}{~~\it{Separate Universe}~~}          \\
     \hline
     \hline
   $~~\pm \epsilon~[{\rm Mpc}/h]^\Delta~~$ & $~~\Delta~~$ & $N_{\rm sim}$ \\ 
    \hline
    0.012 &  0 &  5  \\
    \hline
    0.01 &  0.5  &  5  \\
    \hline
     0.001 &  1  &  5  \\
     \hline
    \end{tabular}
    
   \caption{Parameter values for the simulations used in this work. \emph{Left}: settings for simulations with the squeezed primordial bispectrum in Eq.~\eqref{eq:squeezed_bk_qsfi}. We run 50 realizations for each value of $\Delta$. For $\Delta=0$, we use the first 50 realizations of the \texttt{LC\_p} simulations from the \textsc{Quijote-PNG} suite~\cite{Coulton:2022rir}.
   \emph{Right}: settings for the separate universe simulations used to compute response functions. For each value of $\Delta,$ we run five realizations with the linear power spectrum in Eq.~\eqref{eq:Pk_lin_modification_SU} with $\pm\epsilon$ reported in the table. For the $\Delta=0$ case, we use the first five realizations of the \texttt{s8\_m} and \texttt{s8\_p} \textsc{Quijote} simulations~\cite{Villaescusa-Navarro:2019bje}.}
   \label{tab:sim_parameters}
\end{table}

\subsubsection{Initial conditions with the collider bispectrum}\label{subsec:PNG_IC_sims}
\noindent We generate initial conditions for the PNG simulations based on the method developed in Ref.~\cite{Scoccimarro:2011pz}. Specifically, we express the primordial potential as
\begin{equation}\label{eq:quadratic_png}
    \Phi(\vk)=\phi(\vk)+f_{\rm NL}^{\Delta}\left[\Psi_{\Delta}(\vk) -\langle \Psi_{\Delta}(\vk)\rangle\right],
\end{equation}
where $\phi(\vk)$ is a Gaussian random field, $f_{\rm NL}^\Delta$ is the amplitude of non-Gaussianity, and $\Psi_\Delta(\vk)$ is an auxiliary field that is quadratic in $\phi(\vk)$. Imposing homogeneity and isotropy, the most general form for $\Psi_\Delta(\vk)$ is
\begin{equation*}
    \Psi_\Delta(\vk)=\int\limits_{\vk_1,\vk_2}(2\pi)^3\delta_D(\vk-\vk_{12})\,K_{\Delta}(\vk_1, \vk_2)\,\phi(\vk_1)\,\phi(\vk_2),
\end{equation*}
where $K_{\Delta}(\vk_1, \vk_2)$ is a coupling kernel whose form is determined by demanding that the three-point function of $\Phi(\vk)$ gives the desired bispectrum at leading order in $f_{\rm NL}^{\Delta}.$  For a given bispectrum, the coupling kernel is not uniquely determined~\cite{Schmidt:2010gw, Scoccimarro:2011pz}. Nevertheless, it is important to choose $K_{\Delta}(\vk_1,\vk_2)$ such that loop do not alter the $k^{n_s-4}$ shape of the large-scale power spectrum. In this work, we use the following kernel:\footnote{In practice, we account for departures from scale invariance by replacing all exponents $\Delta \rightarrow \frac{\Delta}{3}(4-n_s)$~\cite{Coulton:2022rir}.}
\begin{equation}\label{eq:kernel_scale_inv}
    K_{\Delta}(\vk_1, \vk_2)=\left[\left(\frac{k_1}{k_{12}}\right)^{\Delta}+\left(\frac{k_2}{k_{12}}\right)^{\Delta}-1\right].
\end{equation}
The squeezed limit ($q\equiv k_1\ll k_2\approx k_{12}\equiv k$) of the bispectrum associated with this kernel is, up to $\mathcal{O}\left(f_{\rm NL}^\Delta\right)$,
\begin{equation}\label{eq:bispectrum_from_kernel}
     \frac{B_{\Phi}(\vq, \vk)}{P_{\Phi}(q)P_{\Phi}(k)}= 4f_{\rm NL}^\Delta \bigg[
     {\left(\frac{q}{k}\right)^\Delta +\left(\frac{k}{q}\right)^\Delta \frac{P_{\Phi}(k)}{P_{\Phi}(q)}}\bigg].
\end{equation}
    If the first term above dominates over the second term, then we recover the squeezed bispectrum in Eq.~\eqref{eq:squeezed_bk_qsfi}.\footnote{We emphasize that Eq.~\eqref{eq:kernel_scale_inv} generates the target \emph{squeezed bispectrum}, which is all that is needed for our work. More sophisticated techniques, (\emph{e.g.},~\cite{Regan:2011zq}) are necessary to simulate non-squeezed bispectrum configurations in the collider scenario. This also requires a template for the full shape dependence of the collider bispectrum, which is both model-dependent and very difficult.} For a scale-invariant primordial power spectrum, the second term is suppressed relative to the first by $((q/k)^{3-2\Delta})$. In this work, we simulate $\Delta=0.5$ and 1; thus we can safely ignore the second contribution.\footnote{For simulations with larger values of $\Delta$, both contributions in Eq.~\eqref{eq:bispectrum_from_kernel} would need to be included in the theoretical model. Both solutions appear in the theoretical model for the Cosmological Collider squeezed bispectrum, where $\Delta'=3-\Delta$ is a second solution to the equations of motion. Indeed, this solution must be included for fields with $m>3H/2$ to ensure the bispectrum is real. However, the amplitude, $f_{\rm NL}^\Delta$, is generally different for the two solutions. Thus, Eq.~\eqref{eq: squeezed_bispectrum_integral} applies only when the second solution is sub-dominant. The exception is for fields with $m=3H/2$ ($\Delta=3/2$), for which Eq.~\eqref{eq:kernel_scale_inv} generates the target bispectrum with an additional factor of two. Alternatively, one could use a high-pass filtered kernel $K_{\Delta}(\vk_1,\vk_2)\rightarrow K_\Delta(\vk_1,\vk_2)\Theta(k_{12}-k_L)$, where $\Theta$ is the step function and $k_L\approx 0.1~h/{\rm Mpc}$. This filter would suppress the second contribution to Eq.~\eqref{eq:bispectrum_from_kernel}.}

Assuming a scale-invariant primordial power spectrum, then the one-loop contribution to $P_{\Phi}$ is
\begin{equation}
    P_{\Phi}^{1-{\rm loop}}(k)\propto \int_{\vp}\left(K_{\Delta}(\vp, \vk-\vp) \right)^2 p^{-3}(|\vk-\vp|)^{-3},
\end{equation}
which is proportional to $k^{-2\Delta}$ at low $k$. For $\Delta< 3/2$, the one-loop corrections are suppressed relative to the linear power spectrum at low $k$. For $\Delta=3/2$, both the $P^{\rm lin.}_{\Phi}$ and $ P_{\Phi}^{1-{\rm loop}}$ scale like $k^{-3};$ however, since the one-loop contribution is proportional to $(f_{\rm NL}^{\Delta})^2$, it is subdominant. Thus, Eq.~\eqref{eq:kernel_scale_inv} does not significantly alter $P_\Phi.$

Finally, we can write the auxiliary field $\Psi_{\Delta}$ in a separable form by replacing the Dirac delta with an exponential,
\begin{align*}
    {{\Psi_\Delta(\vk)}}=\int {d^3{\bx}}\,e^{i\vk\cdot \bx}\bigg[\frac{2\,\phi(\bx)}{k^\Delta}\int_{\vp}{p}^{\Delta}\phi(\vp)e^{-i\vp\cdot \bx}-\phi^2(\bx)\bigg],
\end{align*}
which can be efficiently evaluated using FFTs. In Appendix~\ref{sec:App_spin_and_osc}, we show how to generalize this kernel to account for fields with spin and $m>3H/2.$

We modify the code 2LPTPNG\footnote{\href{https://cosmo.nyu.edu/roman/2LPT/}{https://cosmo.nyu.edu/roman/2LPT/}} to generate initial conditions for the collider bispectrum. We run 50 simulations with $f_{\rm NL}^{\Delta=0.5}=300$ and $f_{\rm NL}^{\Delta=1}=1000$. To isolate the impact of PNG, we use the same seeds as the first 50 realizations of the \textsc{Quijote} and \textsc{Quijote-PNG} simulations. For comparison, we also analyze the same 50 realizations of the fiducial \textsc{Quijote} simulations, which have Gaussian initial conditions, as well as the \texttt{LC\_p} \textsc{Quijote-PNG} simulations, which have $f_{\rm NL}^{\rm loc.}=100.$

\subsubsection{Separate Universe Simulations}\label{subsec:separate_universe}

\noindent We also run separate universe simulations to compute the potential derivative (Eq.~\ref{eq:potential_derivative}) and the non-Gaussian bias (Eq.~\ref{eq:b_psi_SU}). As described in Sec.~\ref{subsec:squeezed_b_primordial_SU}, these derivatives can be estimated numerically from the output of simulations with initial linear power spectra rescaled by $P(k)\rightarrow (1+2\epsilon k^{-\Delta})P(k)$ for small values of $\pm \epsilon$.\footnote{Some initial condition codes recompute the normalization of the input power spectrum to match the input value of $\sigma_8$. In this case, one should fix the normalization to the value computed from the fiducial, \emph{i.e.}, $\epsilon=0$ power spectrum.} We run five realizations for each value of $\pm\epsilon$, which we find sufficient to reduce cosmic variance for the small-scale modes analyzed here. We list our values of $\epsilon$ in Table~\ref{tab:sim_parameters}. For the simulations with $f_{\rm NL}^{\rm loc.}$, \emph{i.e.}, $\Delta=0$, Eq.~\eqref{eq:Pk_lin_modification_SU} is equivalent to a change in the amplitude of the linear power spectrum. Hence, we use the \texttt{s8\_m} and \texttt{s8\_p} simulations from the \textsc{Quijote} suite, which are run with $\sigma_8=0.819$ and $0.834$, respectively.

\subsection{Measurements and likelihood}\label{subsec:measurements_modelling_likelihood}

\noindent In this section, we describe the measurements and likelihood used to study the squeezed matter bispectrum, collapsed matter trispectrum, and scale-dependent halo bias for the squeezed bispectrum in Eq.~\eqref{eq:squeezed_bk_qsfi} with $N$-body simulations. Our method is based on the bispectrum analysis presented in \citet{Goldstein:2022hgr}. We refer the reader there for a more detailed description of our modeling and measurement choices. We use the \texttt{PNGolin} package\footnote{\href{https://github.com/samgolds/PNGolin}{https://github.com/samgolds/PNGolin}} to compute the power spectrum, bispectrum, and trispectrum from the simulation output.  See Appendix~\ref{sec:App_estimators} for a detailed discussion of our estimators.

\subsubsection{Measurements}

\noindent We aim to study imprints of the Cosmological Collider on higher-order statistics of the matter field, as well as the two-point statistics of the halo field. To estimate these summary statistics, we first assign the redshift $z=0$ matter particles and halos to a $1024^3$ mesh using Cloud-in-Cell interpolation with the \textsc{Pylians} package~\cite{Pylians}.\footnote{\href{https://pylians3.readthedocs.io/en/master/}{https://pylians3.readthedocs.io/en/master/}} We split the FoF halo catalog into four logarithmically spaced mass bins with $13.5<\log_{10}(M_{\rm FoF})<14.5~M_{\odot}/h$, as well as one bin with $M_{\rm FoF}>10^{14.5}~M_\odot/h$. From these fields, we compute the matter power spectrum, halo power spectrum, and halo-matter cross-spectrum in bins of width $2k_f$, where $k_f\equiv 2\pi/L_{\rm box}\approx 0.006~h/{\rm Mpc}$ is the fundamental mode. We always exclude the fundamental mode from our analyses to avoid numerical issues.

We analyze the squeezed bispectrum integrated over a range of hard modes. Specifically, we compute the squeezed bispectrum for soft modes $q>k_f$ in bins of width $2k_f$ up to $q_{\rm max}=0.05~h/{\rm Mpc}$ and averaged over hard modes in some bin $k_{\rm min}<k,k'<k_{\rm max}$.\footnote{This definition differs slightly from the bispectrum estimator used in Ref.~\cite{Goldstein:2022hgr}, which required $k_{\rm min}<k<k_{\rm max}$ and let $k'$ take any value allowed by the triangle inequality. Not imposing the condition $k_{\rm min}<k'<k_{\rm max}$ leads to a somewhat enhanced $q$ dependence in the squeezed bispectrum.} This corresponds to $N_q=7$ soft mode bins for each choice of $k_{\rm min}$ and $k_{\rm max}$. We use $B_{m}(q|\,\mathcal{K})$ to denote the integrated squeezed bispectrum as a function of the soft mode, $q$, averaged over \emph{all} hard modes $k,k'\in\mathcal{K}$. In Sec.~\ref{sec:info_content}, we compare the information content of the squeezed bispectrum integrated over a single bin $\mathcal{K}$ with the squeezed bispectrum split into multiple disjoint sub-bins, \emph{e.g.}, $B_m(q|\,\mathcal{K}_a)$ and $B_m(q|\,\mathcal{K}_b)$ where $\mathcal{K}_a\cup \mathcal{K}_b=\mathcal{K}.$

Similar to the squeezed bispectrum measurements, we compute the collapsed trispectrum for soft modes $q\equiv k_{12}>k_f$ in bins of width $2k_f$ up to $q_{\rm max}=0.05~h/{\rm Mpc}$ and averaged over all hard modes in some bin $k_{\rm min}<k_1,k_2,k_3,k_4<k_{\rm max}.$ We use the notation $T_m(q|\,\mathcal{K})$ to denote the integrated trispectrum as a function of the soft mode, $q$, averaged over all hard modes $k_1,k_2,k_3,k_4\in\mathcal{K}$. We also analyze the collapsed trispectrum divided into two hard mode bins, $\mathcal{K}_a$ and $\mathcal{K}_b$. In this case, there are three configurations $T_m(q|\,\mathcal{K}_a),$ $T_m(q|\,\mathcal{K}_b)$, and $T_m(q|\,\mathcal{K}_a, \mathcal{K}_b)$, where the final configuration corresponds to the asymmetric configuration with $k_1,k_2\in \mathcal{K}_a$ and $k_3,k_4 \in \mathcal{K}_b$ (see Fig.~\ref{fig:squeezed_B_collapsed_T_shape_dependence}).

To avoid binning effects due to the finite number of modes at low $q$ in our simulations, we explicitly bin our theory model by computing $\langle q^{-2+\Delta}\delta_m(\vq) \delta_m(-\vq)/T(q)\rangle$ and $\langle q^2 \delta_m(\vq)\delta_m(-\vq)\rangle$ using the discrete density fields. This is described in detail in~\citet{Goldstein:2022hgr}.

\subsubsection{Likelihood}

\noindent We now describe how we fit our simulation measurements to constrain the amplitude of PNG, $f_{\rm NL}^\Delta$. We focus on the likelihood for the squeezed matter bispectrum in a single hard mode bin, which includes all of the main ingredients without the notational complexity. We explain how to generalize this to the other summary statistics considered in this work at the end of this section.

Let ${\bm{\theta}}$ denote the vector of parameters that we vary in a fit, \emph{i.e.}, $f_{\rm NL}^\Delta$, $\bar{a}_0$, and $\bar{a}_2$ for the bispectrum in a single bin. Since the main goal of this work is to demonstrate that our non-perturbative models for the squeezed matter bispectrum and collapsed trispectrum are unbiased, we often use a joint likelihood in the soft matter power spectrum, $P_m(q)$, which reduces cosmic variance. As described in~\citet{Goldstein:2022hgr}, the joint likelihood for $B_m(q|\,\mathcal{K})$ and $P_m(q)$ can be written as a likelihood in $B_m(q|\,\mathcal{K})$ with a parameter-dependent covariance that depends on $P_m(q)$.\footnote{In Sec.~\ref{sec:info_content}, we show constraints from a likelihood in just $B_m(q|\,\mathcal{K})$. We determine these constraints from the same procedure described in this section, \emph{except} we compute $P_m(q)$ from linear theory and we do not use a parameter dependent covariance.  } Specifically, 
\begin{align} \label{eq:likelihood}
    \mathcal{L} \propto \frac{1}{\sqrt{\det {\mathcal{C}}({\bm{\theta}})}}\bigg[1+\frac{\delta B({\bm{\theta}})\cdot{\mathcal{C}}^{-1}({\bm{\theta}})\cdot\delta  B({\bm{\theta}})}{N_{\rm sim}-1}\bigg]^{-\frac{N_{\rm sim}}{2}} \,,
\end{align}
where $\delta B({\bm{\theta}}) \equiv {B}_m(q|\,\mathcal{K})-B_{\rm thr.}(q, \mathcal{K}|{\bm{\theta}})$ is the difference between the measured bispectrum and our theory model (Eq.~\ref{eq: squeezed_bispectrum_integral}), 
${\mathcal{C}}_{ij}({\bm{\theta}})\equiv \langle\delta B_i({\bm{\theta}})\delta B_j({\bm{\theta}})\rangle$ is the \emph{parameter-dependent} covariance of $\delta B$, and $N_{\rm sim}$ is the number of realizations used to estimate it. We estimate the covariance using $N_{\rm sim}=50$ simulation realizations. To mitigate numerical instabilities, we assume that the covariance is diagonal in $q$, which is a good approximation over the scales we consider~\cite{Giri:2023mpg}. Finally, we use a multivariate $t$-distribution to account for the fact that our covariance is estimated from a finite number of simulations~\cite{Sellentin:2015waz}.

\begin{figure*}[!t]
\centering
\includegraphics[width=0.995\linewidth]{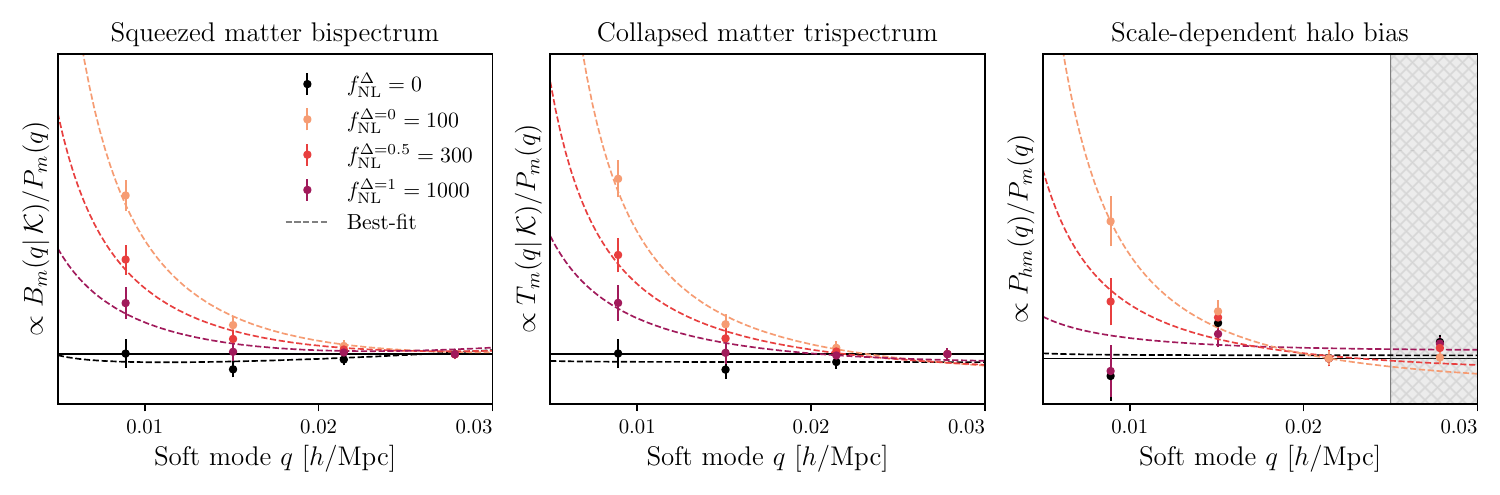}
\caption{Impact of intermediate-mass scalars during inflation on the soft limits of large-scale structure correlation functions. The solid points show redshift $z=0$ measurements of the squeezed matter bispectrum (left), collapsed matter trispectrum (middle), and scale-dependent halo bias (right) from $N$-body simulations for different values of the amplitude ($f_{\rm NL}^\Delta$) and particle mass ($\Delta$).  To highlight the different scalings, we divide all measurements by the power spectrum, $P_m(q)$, and fix the ratios to one at $q=0.028~h/{\rm Mpc}$ for the bispectrum and trispectrum and $q=0.021~h/{\rm Mpc}$ for the scale-dependent bias. The squeezed primordial bispectrum in Eq.~\eqref{eq:squeezed_bk_qsfi} leads to poles whose amplitude and shape depend on $f_{\rm NL}^\Delta$ and $\Delta.$ The dashed lines show our best-fit theory models given by Eq.~\eqref{eq:squeezed_bispectrum_non_pert}, Eq.~\eqref{eq:collapsed_T_no_contact}, and Eq.~\eqref{eq:Phm}, for the bispectrum, trispectrum, and the scale-dependent bias, respectively. The shaded grey region indicates modes excluded from the halo analysis due to non-linear biasing. See the main text for precise definitions of $B_m(q|\,\mathcal{K})$ and $T_m(q|\,\mathcal{K}),$ as well as the scale cuts and halo masses used in the analysis shown here.}
\label{fig:measurements_with_fit}
\end{figure*}

Since sampling from a parameter-dependent covariance is costly, we evaluate the covariance with the parameters fixed to their \emph{maximum a posteriori} values using the iterative maximization procedure described in Ref.~\cite{Goldstein:2022hgr}. In Ref.~\cite{Goldstein:2022hgr}, we found that this procedure has negligible impact on our results. With the covariance determined, we sample from the parameter posterior using the \texttt{emcee} package~\cite{Foreman-Mackey:2012any} assuming uniform wide priors with $-10^5<f_{\rm NL}^\Delta, \bar{a}_i<10^5$. We assess convergence using the Gelman-Rubin statistic~\cite{gelman1992} with a
tolerance of $|R-1|<0.01$.

To generalize the above discussion to a likelihood in the squeezed bispectrum measured in $N_k$ hard mode bins, $\mathcal{K}_1,\dots,\mathcal{K}_n$, we replace the $N_{q}$-dimensional bispectrum data-vector with an $N_q\times N_k$-dimensional data-vector of bispectrum measurements, $B(q|\mathcal{K}_i)$, in each hard mode bin. We estimate the covariance from simulations assuming it is diagonal in $q$ but including off-diagonal correlations between different $k$ bins. For each $k$ bin, we introduce a new set of free coefficients $\bar{a}_0^{(i)}$ and $\bar{a}_2^{(i)}$; thus we have $2N_k+1$ free parameters for the bispectrum model. A similar approach was recently used in Ref.~\cite{Giri:2023mpg}. Here, we analyze the squeezed bispectrum for $N_k=1,$ $2$, and $4$.

The collapsed trispectrum likelihood in a single hard mode bin takes the same form as the single-bin bispectrum likelihood with $\delta B({\bm{\theta}})\rightarrow \delta T_m({\bm{\theta}})=T_m(q|\mathcal{K})-T_{\rm thr.}(q,\mathcal{K}|{\bm{\theta}}).$ In addition to the free parameters in the bispectrum model, the trispectrum theory model includes $\bar{T}_0$. We vary $\bar{T}_0$ in our trispectrum analyses assuming a uniform prior $5<\log_{10}\left(\bar{T}_0~({\rm Mpc}/h)^9\right)<15$. For the simulation volume and scale cuts used in this work, we find that it is sufficient to fix $\bar{a}_2=0$, while varying $f_{\rm NL}^\Delta, \bar{a}_0$, $\bar{T}_0$.\footnote{If we vary $\bar{a}_2$, then our constraints on $f_{\rm NL}^\Delta$ are still unbiased. However, the error on $f_{\rm NL}^\Delta$ increases significantly because of the degeneracy between $\bar{a}_2$ and $\bar{T}_0$.} Thus, we fix $\bar{a}_2=0$ for all trispectrum analyses in this work, except for the joint analyses of the squeezed bispectrum and collapsed trispectrum, as including the squeezed bispectrum breaks the $\bar{T}_0-\bar{a}_2$ degeneracy. 

We can generalize the collapsed trispectrum likelihood to include measurements in $N_k$ hard mode bins. This procedure is identical to the generalization to the multi-bin bispectrum likelihood, except the data vector also includes the $\binom{N_k}{2}$ asymmetric trispectrum configurations, $T(q|\mathcal{K}_i, \mathcal{K}_j).$ Importantly, $\log_{10}(\bar{T}_{i,j})$ is the only free parameter that needs to be added to analyze the asymmetric configurations. Here, we analyze the collapsed trispectrum with $N_k=1$ and $N_k=2$. Finally, we can jointly analyze the squeezed matter bispectrum and collapsed trispectrum in multiple hard mode bins by combining the multi-bin bispectrum and trispectrum likelihoods, including the covariance between the squeezed bispectrum and collapsed trispectrum at a fixed soft mode, $q.$ We fit the squeezed matter bispectrum and collapsed trispectrum up to $q_{\rm max}=0.05~h/{\rm Mpc}.$

We study the scale-dependent halo bias using a joint likelihood in $P_{h}(q)$, $P_{hm}(q)$, and $P_{m}(q)$ to reduce cosmic variance. We fix $f_{\rm NL}^\Delta$ to its true value for a given simulation while varying the linear bias ($b_1$), the non-Gaussian bias ($b_{\Psi,\Delta}$), and the stochastic contribution ($\epsilon_N$). We assume the following uniform priors: $0<b_1<5$, $-20<b_{\Psi,\Delta}<20$, and $-1<\epsilon_N<1$. We fit the halo statistics up to $q_{\rm max} = 0.025~h/{\rm Mpc}$, beyond which higher-order bias terms are required given the high-mass sample and large $f_{\rm NL}^{\rm loc.}$ value.

\section{Results}\label{sec:results}

\noindent Fig.~\ref{fig:measurements_with_fit} is one of the main results of this work: comparison of our squeezed limit $N$-point function models to measurements from numerical simulations. Using simulations with and without PNG, we plot measurements of the ratio of the squeezed matter bispectrum (left), collapsed matter trispectrum (center), and halo-matter cross-spectrum (right) relative to the soft matter power spectrum $P_{m}(q)$ (all at redshift zero). For simulations without PNG, all ratios approach a constant on large scales. On the other hand, for simulations with PNG described by the squeezed bispectrum in Eq.~\eqref{eq:squeezed_bk_qsfi}, these ratios have poles in the soft limits. The amplitude and shape of these poles are set by $f_{\rm NL}^\Delta$ and $\Delta$, respectively.  

The dashed lines show the best-fit theory models given by Eq.~\eqref{eq:squeezed_bispectrum_non_pert}, Eq.~\eqref{eq:collapsed_T_no_contact}, and Eq.~\eqref{eq:Phm}, for the squeezed matter bispectrum, collapsed matter trispectrum, and the scale-dependent bias, respectively. For the matter bispectrum and trispectrum, we determine the best-fit model parameters from a joint analysis of the soft matter power spectrum, squeezed matter bispectrum, and collapsed matter trispectrum with soft modes between $0.006<q<0.05~h/{\rm Mpc}$ and hard modes averaged over a single bin with $0.4<k<1.1~h/{\rm Mpc}$. For the scale-dependent halo bias, we determine the best-fit model parameters from a joint analysis of the halo power spectrum, halo-matter cross-spectrum, and matter power spectrum for halos with mass between ${14.25}<\log_{10}(M_{\rm FoF})<14.5~M_\odot/h$. The scale-dependent bias analysis excludes modes with $q>0.025~h/{\rm Mpc}$ because the linear biasing model breaks down beyond these scales. The best-fit theory predictions provide an excellent fit to the data for all summary statistics and simulations considered, thus validating the models discussed above.

\begin{figure}[!t]
\centering
\includegraphics[width=0.99\linewidth]{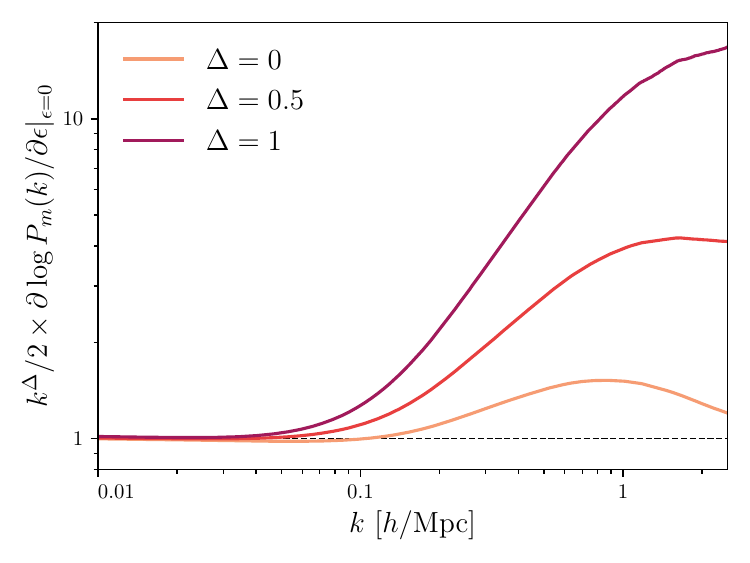}
\caption{Non-linear contribution to the potential derivative (Eq.~\ref{eq:potential_derivative}) estimated from separate universe simulations for the three values of $\Delta$ analyzed in this work. The derivative is computed at redshift $z=0$. On large scales, the potential derivative agrees with linear theory. On small scales, the derivative is amplified by non-linear structure formation, particularly for large values of $\Delta.$ }
\label{fig:response_function}
\end{figure}
\subsection{Validation of matter bispectrum and trispectrum models}

\noindent To obtain constraints on $f_{\rm NL}^{\Delta}$ from the squeezed matter bispectrum and collapsed matter trispectrum, we first need to estimate the non-linear potential derivative in Eq.~\eqref{eq:potential_derivative} from the separate universe simulations. Fig.~\ref{fig:response_function} shows the non-linear enhancement of the potential derivative as a function of the wave-number for the three values of $\Delta$ analyzed in this work. The derivatives are computed numerically using the method described in Sec.~\ref{subsec:separate_universe}. We average the potential derivative over five realizations to minimize cosmic variance. On large scales, the ratio $\partial \log P_m(k)/\partial \epsilon|_{\epsilon=0}/(2/k^\Delta)$ approaches unity, as expected from linear theory. The small-scale potential derivative is significantly enhanced due to non-linear structure formation, particularly for large values of $\Delta.$ This suggests that the mode coupling induced by the primordial bispectrum in Eq.~\eqref{eq:squeezed_bk_qsfi} could significantly impact small-scale structures for large values of $\Delta.$ However, a detailed investigation of the shape of the potential derivative is beyond the scope of this work.

\begin{figure*}[!t]
\centering
\includegraphics[width=0.995\linewidth]{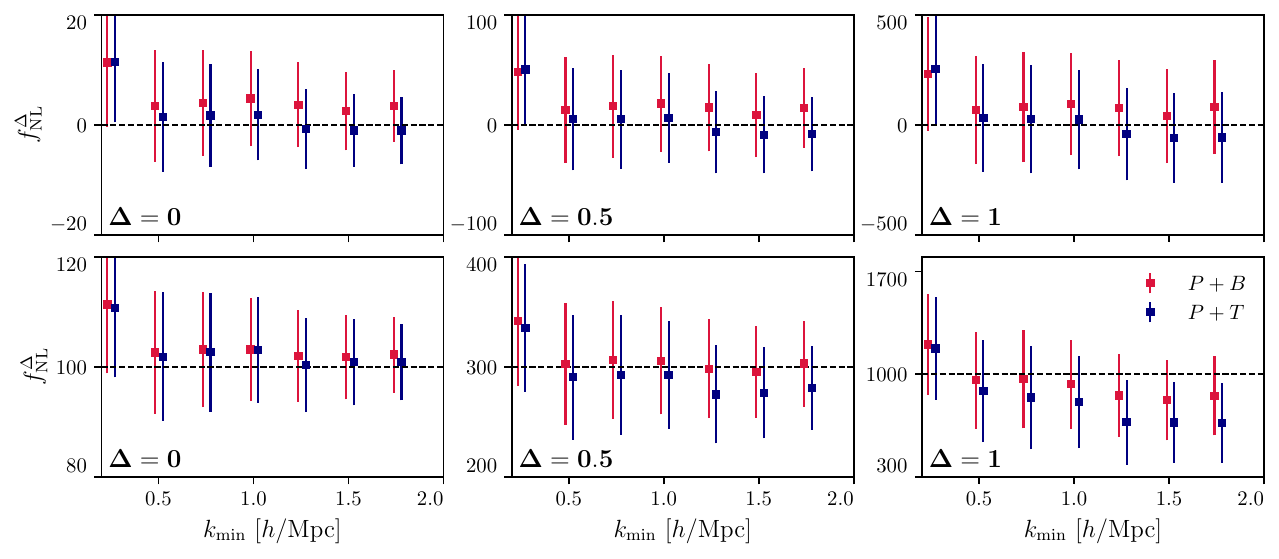}
\caption{Mean and 68\% confidence interval on the inferred value of $f_{\rm NL}^\Delta$ from a joint analysis of the squeezed matter bispectrum and power spectrum (red), as well as the collapsed matter trispectrum and power spectrum (blue) as a function of the \emph{minimum} hard mode, $k_{\rm min}.$ For Gaussian initial conditions (top), we correctly infer $f_{\rm NL}^\Delta=0$ from fits to the bispectrum and trispectrum measured over a wide range of non-linear modes. The error on $f_{\rm NL}^\Delta$ increases significantly for larger values of $\Delta$. We also find unbiased constraints when analyzing simulations with PNG described by the squeezed primordial bispectrum in Eq.~\eqref{eq:squeezed_bk_qsfi} (bottom). This demonstrates that the non-perturbative models for the squeezed matter bispectrum and collapsed matter trispectrum derived in this work can be used to constrain the Cosmological Collider deep into the non-linear regime. Due to the non-Gaussian covariance, the constraints are highly correlated across different $k$ bins.}
\label{fig:fits_vary_kmin}
\end{figure*}

Equipped with the separate universe potential derivatives and measurements of the squeezed matter bispectrum and collapsed matter trispectrum from $N$-body simulations, we turn to the main task: constraining $f_{\rm NL}^\Delta$. Fig.~\ref{fig:fits_vary_kmin} shows the mean value of $f_{\rm NL}^\Delta$ and its 68\% confidence interval as a function of the minimum hard momentum, $k_{\rm min}$, inferred from the squeezed bispectrum (red) and collapsed trispectrum (blue) for the simulations considered in this work. The horizontal dashed lines indicate the true value of $f_{\rm NL}^\Delta$. For each value of $k_{\rm min},$ we integrate over all hard modes in a narrow bin with $k_{\rm max}=k_{\rm min}+0.06~h/{\rm Mpc}$. This allows us to robustly test the scale dependence of the response functions estimated from the separate universe simulations. Notably, any incorrect prediction of this scale dependence would significantly bias the constraint on $f_{\rm NL}^\Delta$ (see Fig.~\ref{fig:response_function}). To assess potential biases in our theoretical models, we use a joint likelihood with the soft power spectrum $P_m(q)$, which minimizes the cosmic variance contribution from the long-wavelength modes. Moreover, we fit the mean data-vector estimated from $N_{\rm sim}=50$ simulations, rescaling the covariance by $1/N_{\rm sim}.$ 

The top panels of Fig.~\ref{fig:fits_vary_kmin} shows the constraints on $f_{\rm NL}^\Delta$ derived from the simulations with Gaussian initial conditions for different values of $\Delta$. In all cases, we recover $f_{\rm NL}^\Delta=0$. The constraints are slightly biased for $k_{\rm min}\approx 0.25~h/{\rm Mpc}$, suggesting that these triangle configurations may not be sufficiently squeezed for our bispectrum and trispectrum models to apply. Notice that the error on $f_{\rm NL}^\Delta$ grows dramatically as $\Delta$ increases. This is expected because $B_m(q|\mathcal{K})\propto f_{\rm NL}^\Delta/q^{2-\Delta}$, meaning that for a fixed $f_{\rm NL}^\Delta$, the signal becomes harder to detect at higher values of $\Delta$.  Furthermore, with larger $\Delta$, $f_{\rm NL}^\Delta$ becomes increasingly degenerate with the nuisance parameter $\bar{a}_0$, resulting in an additional penalty when marginalizing over $\bar{a}_0$. We discuss this in more detail in Sec.~\ref{sec:info_content}.

The bottom panels of Fig.~\ref{fig:fits_vary_kmin} presents the constraints on $f_{\rm NL}^\Delta$ derived from the simulations with PNG. In all cases, we correctly infer the amplitude $f_{\rm NL}^\Delta$ up to scales of $2~h/{\rm Mpc}$.\footnote{In principle, we can go beyond $k=2~h/{\rm Mpc}$. However, given the numerical resolution of our simulations, we choose to analyze only modes with $k<2~h/{\rm Mpc}$. Using similar methods, \citet{Giri:2023mpg} constrained $f_{\rm NL}^{\rm loc.}$ with the squeezed bispectrum and collapsed trispectrum from the \textsc{Quijote} simulations up to $k=3~h/{\rm Mpc}.$} It is worth noting that the constraints from the trispectrum for the $\Delta=1$ simulations are slightly low at $k_{\rm min}\gtrsim 1.2~h/{\rm Mpc}$. This could be a statistical fluctuation, or, alternatively, could arise since we fix $\bar{a}_2=0$ in the trispectrum model thus neglecting higher-order ($q^2/k^2$) contributions. Finally, for the $\Delta=1$ simulations, the loop contributions to the matter power spectrum are $\mathcal{O}(15\%)$, which can lead to slight biases in the constraints.\footnote{By eye, one may worry that all of the constraints in Fig.~\ref{fig:fits_vary_kmin} look suspiciously good. However, we stress that due to the non-Gaussian squeezed bispectrum and collapsed trispectrum covariances, the inferred value of $f_{\rm NL}^\Delta$ is highly correlated as a function of $k_{\rm min}$. Furthermore, since all simulations are run using the same seeds, the residuals will be correlated \emph{across} simulations. Finally, rescaling the sample covariance by $1/N_{\rm sim}$ is only an approximate measure of the covariance of the mean.} 

In conclusion, given the significant dependence of the potential derivative, $\partial P_m(k|\epsilon, \Delta)/\partial \epsilon$, on both $\Delta$ and $k$ (see Fig.~\ref{fig:response_function}), the results of this section further validate our separate universe models for the squeezed matter bispectrum and collapsed matter trispectrum.

\subsection{Information content of the matter bispectrum and trispectrum}\label{sec:info_content}

\noindent Having demonstrated that our non-perturbative models for the squeezed matter bispectrum and collapsed matter trispectrum yield unbiased constraints on $f_{\rm NL}^\Delta$ deep into the non-linear regime, we now assess the information content of these summary statistics. All results in this section use the mean bispectrum and trispectrum measurements averaged over 50 realizations and the covariance estimated for a single realization with a  $(1~{\rm Gpc}/h)^3$ volume. We first focus on the squeezed bispectrum.

\begin{figure}[!t]
\centering
\includegraphics[width=0.995\linewidth]{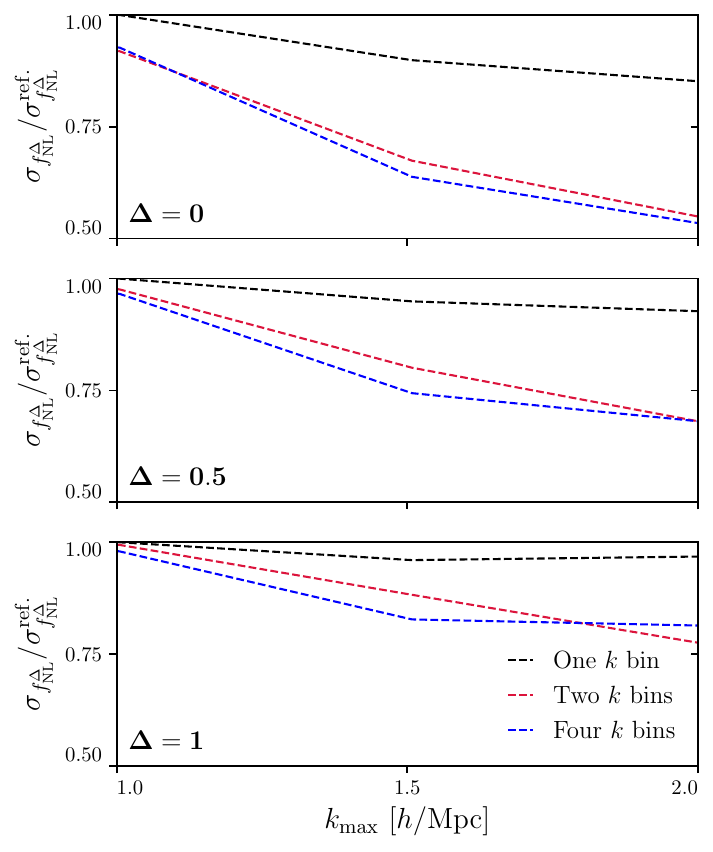}
\caption{Dependence of the 68\% error on $f_{\rm NL}^\Delta$ on the maximum hard mode, $k_{\rm max}$, and the number of $k$ bins for different values of $\Delta$. For the single-bin analysis (black), increasing $k_{\rm max}$ yields minimal improvement. In contrast, multi-bin analyses (red and blue) show significant improvement at higher $k_{\rm max}$ due to sample variance cancellation (see the main text for details). For larger $\Delta$, $\sigma_{f_{\rm NL}^\Delta}$ is less sensitive to $k_{\rm max}$ because of the additional $k^{-\Delta}$ scaling of the primordial squeezed bispectrum (Eq.~\ref{eq:squeezed_bk_qsfi}). All errors are estimated from a joint analysis of the redshift $z=0$ squeezed matter bispectrum and large-scale matter power spectrum, assuming $k_{\rm min}=0.38~h/{\rm Mpc}$. }
\label{fig:fits_function_kmax}
\end{figure}

In Fig.~\ref{fig:fits_function_kmax}, we plot the $1\sigma$ error on $f_{\rm NL}^\Delta$ as a function of the maximum hard mode, $k_{\rm max}$, from a joint analysis of the squeezed matter bispectrum and large-scale matter power spectrum for three different values of $\Delta$. We analyze the squeezed bispectrum with soft modes between $0.006<q<0.05~h/{\rm Mpc}$ and hard modes between $0.38<k<k_{\rm max}~h/{\rm Mpc}$. We consider three values of $k_{\rm max}$ between 1 and $2~h/{\rm Mpc}.$ For each value of $k_{\rm max}$, we show the error for an analysis with one $k$ bin (black), two $k$ bins (red), and four $k$ bins (blue). To emphasize how the constraints depend on the number of $k$ bins and choice of $k_{\rm max}$, we divide all errors by the error from the single bin analyses at $k_{\rm max}=1~h/{\rm Mpc}.$ The different panels correspond to different values of $\Delta$. All errors are derived by fitting the simulations with Gaussian initial conditions.

For the single bin analysis, the constraints improve only marginally when increasing $k_{\rm max}$. Indeed, the most significant improvement is seen in the local PNG analysis ($\Delta=0$), where the constraint on $f_{\rm NL}^\Delta$ improves by 15\% after increasing $k_{\rm max}$ from $1$ to $2~h/{\rm Mpc}$. This is consistent with Refs.~\cite{Goldstein:2022hgr, Giri:2023mpg}, which found that the constraints on $f_{\rm NL}^{\rm loc.}$ from the squeezed matter bispectrum averaged over a single bin saturate quickly as a function of $k_{\rm max}$ due to the non-Gaussian covariance~\cite{Biagetti:2021tua, Floss:2022wkq}. 

\begin{figure}[!t]
\centering
\includegraphics[width=0.92\linewidth]{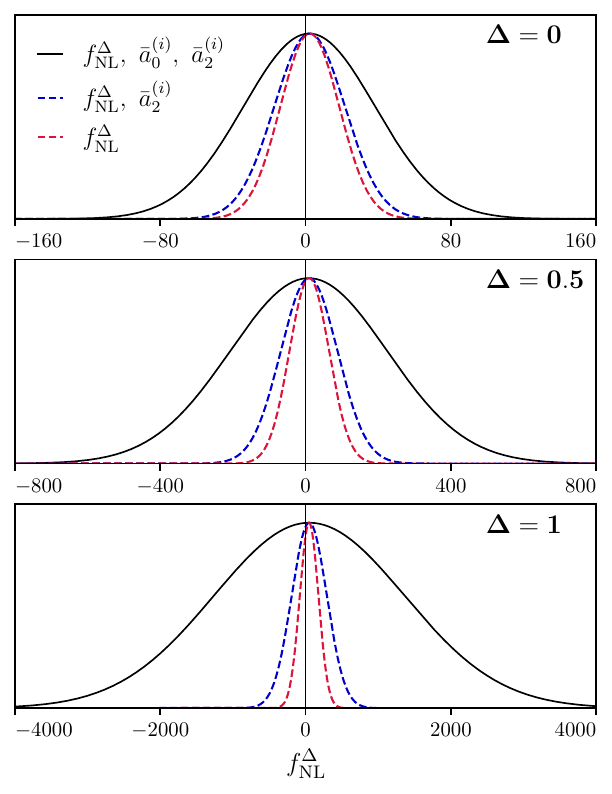}
\caption{Marginalized posterior on $f_{\rm NL}^\Delta$ for a joint analysis of the squeezed matter bispectrum and soft matter power spectrum with different assumptions about the nuisance parameters, $a_{0}^{(i)}$ and $a_{2}^{(i)}$, describing late-time physics. The error on $f_{\rm NL}^\Delta$ is significantly larger for high values of $\Delta$. This is primarily due to the degeneracy between $f_{\rm NL}^\Delta$ and $\bar{a}_0^{(i)}$. For an \emph{idealistic} analysis with fixed $a_{0}^{(i)}$, the error on $f_{\rm NL}^\Delta$ improves by a factor of 2, 3, and 5 for $\Delta=0,~0.5,$ and $1$, respectively. Fixing $a_{2}^{(i)}$ leads to further improvements. Table~\ref{tab:sigma_fNL} lists the 68\% confidence error bar for these analyses.}
\label{fig:marg_bias}
\end{figure}
\begin{table}[!t]
    \centering
  \begin{tabular}{ |c|c|c|c| } 
     \multicolumn{4}{c}{\textbf{{Marginalized error on $\bm{f_{\rm NL}^\Delta}$} }}          \\
     \hline
     \hline
  $~~\Delta~~$ & ~~Fiducial~~  & ~~Fixed $a_{0}^{(i)}$~~ & ~~Fixed $a_{0}^{(i)}$ and $a_2^{(i)}$~~  \\ 
    \hline
     0 &  36  & 19 & 16\\
    \hline
    0.5  &  217 & 78  & 54\\
    \hline
     1  &  1300 & 243 & 128 \\
     \hline
    \end{tabular}
    \
    
   \caption{68\% confidence error bar on $f_{\rm NL}^\Delta$ from a joint analysis of the squeezed matter bispectrum and matter power spectrum from a $(1~{\rm Gpc}/h)^3$ volume with different assumptions about the nuisance parameters, $a_{0}^{(i)}$ and $a_{2}^{(i)}$. The squeezed bispectrum is measured with $0.006<q<0.05~h/{\rm Mpc}$ and averaged over two equal-width hard mode bins with $0.38<k<2.01~h/{\rm Mpc}$. For large values of $\Delta$, marginalizing over $a_{0}^{(i)}$ and $a_{2}^{(i)}$ significantly degrades the constraint on $f_{\rm NL}^\Delta.$ }
   \label{tab:sigma_fNL}
\end{table}

When dividing the bispectrum into multiple $k$ bins (red and blue), the error on $f_{\rm NL}^\Delta$ is notably more sensitive to $k_{\rm max}$. For $\Delta=0$, we find an improvement of nearly 40\% going from $k_{\rm max}=1$ to $2~h/{\rm Mpc}$, consistent with previous works~\cite{Giri:2023mpg}. This improvement arises because, as discussed extensively in Ref.~\cite{dePutter:2018jqk}, analyzing the bispectrum in multiple $k$ bins cancels the cosmic variance associated with the long mode in the bispectrum, $\delta_m(\vq)$. Physically, this cosmic variance cancellation is analogous to the cancellation that occurs when constraining $f_{\rm NL}^{\rm loc.}$ using the scale-dependent bias in a multi-tracer survey~\cite{Seljak:2008xr, McDonald:2008sh, Hamaus:2012ap, Ferraro:2014jba}. In our case, the locally measured small-scale matter power spectrum in multiple different $k$-bins act as multiple biased tracers of the long-wavelength mode. Note that we find negligible further improvement if we use four bins instead of two.

From Fig.~\ref{fig:fits_function_kmax}, it is evident that for larger values of $\Delta,$ the constraints are less sensitive to the choice of $k_{\rm max}$. In linear theory, this is expected since the bispectrum is suppressed by the $k^{-\Delta}$ scaling. However, this argument does not necessarily hold in the non-linear regime, where the $k$-dependence of the bispectrum is described by the potential derivative, $\partial P_m(k|\epsilon, \Delta)/\partial \epsilon$. Indeed, Fig.~\ref{fig:response_function} shows that while the non-linear contribution to the logarithmic derivative decreases as a function of $k$ over $1<k<2~h/{\rm Mpc}$ for the $\Delta=0$ simulations, it \emph{increases} over these scales for the $\Delta=1$ simulations. Nevertheless, this increase is insufficient to offset the steep $k^{-\Delta}$ contribution. Thus, the $\Delta=0.5$ and $\Delta=1$ squeezed bispectra are still suppressed relative to the $\Delta=0$ simulations over the range of hard modes that we analyze here.

Having analyzed the dependence of our constraints on $k_{\rm max}$ and the number of hard mode bins, we now study how $\sigma_{f_{\rm NL}^\Delta}$ depends on the value of $\Delta.$ Fig.~\ref{fig:marg_bias} shows the marginalized posterior on $f_{\rm NL}^\Delta$ from a joint analysis of the squeezed matter bispectrum and large-scale matter power spectrum for the three different values of $\Delta$ considered in this work. We use the bispectrum estimated in two equal-width bins with $k_{\rm min}=60k_f\approx 0.38~h/{\rm Mpc}$ and $k_{\rm max}=320k_f\approx 2.01~h/{\rm Mpc}.$ We show constraints from our fiducial analysis where we vary the coefficients, $\bar{a}_0^{(i)}$ and $\bar{a}_2^{(i)}$ (black), as well as two \emph{idealistic} analyses where we fix $\bar{a}_0^{(i)}$ (blue) and $\bar{a}_0^{(i)}$ and $\bar{a}_2^{(i)}$ (red), practically assuming perfect knowledge of structure-formation physics. Table~\ref{tab:sigma_fNL} lists the 68\% confidence error bar for these analyses.

\begin{figure}[!t]
\centering
\includegraphics[width=0.99\linewidth]{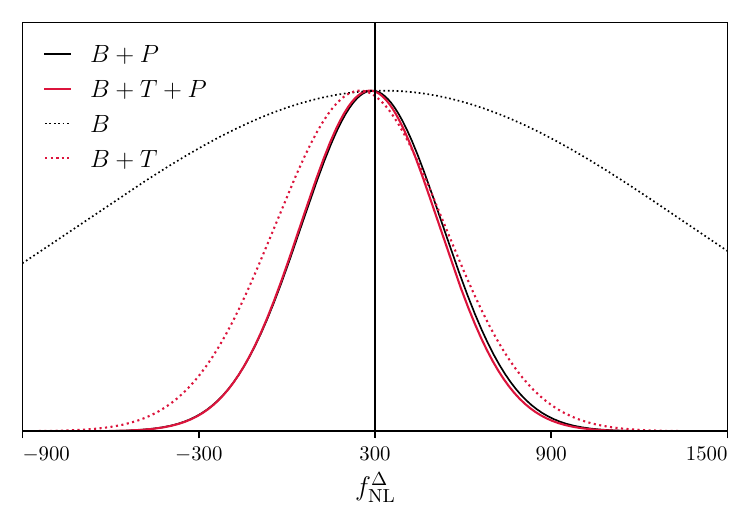}
\caption{Constraint on $f_{\rm NL}^\Delta$ from the squeezed bispectrum (blue) vs. a joint analysis of the squeezed bispectrum and collapsed trispectrum (red). If the particular realization of the matter power spectrum is known, the collapsed trispectrum does not improve the constraint on $f_{\rm NL}^\Delta$ (solid lines). Conversely, the collapsed trispectrum significantly improves constraints if the particular realization of $P_m(q)$ is unknown because it helps cancel cosmic variance.} \label{fig:trispectrum_SV_cancellation}
\end{figure}

In the fiducial analysis, the error on $f_{\rm NL}^\Delta$ is over 36 times larger for $\Delta=1$ compared to $\Delta=0$. This significant difference in constraining power arises from two main factors: (i) the milder scaling of the squeezed-bispectrum pole for large values of $\Delta$, and (ii) the increased degeneracy between $f_{\rm NL}^\Delta$ and our model's nuisance parameters for large values of $\Delta$. To assess how much information is lost upon marginalizing over $\bar{a}_0^{(i)}$ and $\bar{a}_2^{(i)}$, we repeat the fiducial analysis with $\bar{a}_0^{(i)}$ and $\bar{a}_2^{(i)}$ fixed to their \emph{maximum a posteriori} values, thus varying only $f_{\rm NL}^\Delta$. While the error bar for the $\Delta=0$ analysis shrinks by a factor of two, the error bar for the $\Delta=1$ analysis decreases by over a factor of ten. We still find significant improvements in the constraints if we fix $a_{0}^{(i)}$, but vary $a_{2}^{(i)}$. This demonstrates that marginalizing over nuisance parameters, particularly $a_{0}^{(i)}$, significantly limits how well we can constrain $f_{\rm NL}^\Delta$ using the method presented here. The same situation arises when trying to constrain these models using the scale-dependent bias, where marginalization over galaxy bias parameters significantly degrades the constraints~\cite{Green:2023uyz}. Therefore, to optimally constrain $f_{\rm NL}^\Delta$ for large values of $\Delta$, it would be extremely valuable to have informative priors for $\bar{a}_0^{(i)}$ and $\bar{a}_2^{(i)}$. While this may be possible for matter using theory~\cite{Valageas:2013zda, Nishimichi:2014jna} or $N$-body simulations \emph{e.g.},~\cite{Chiang:2017jnm, Biagetti:2022ckz}, it is significantly more challenging for galaxies due to uncertainties in galaxy formation (however, see Ref.~\cite{Ivanov:2024hgq} for recent progress on this front).

\begin{figure}[!t]
\centering
\includegraphics[width=0.99\linewidth]{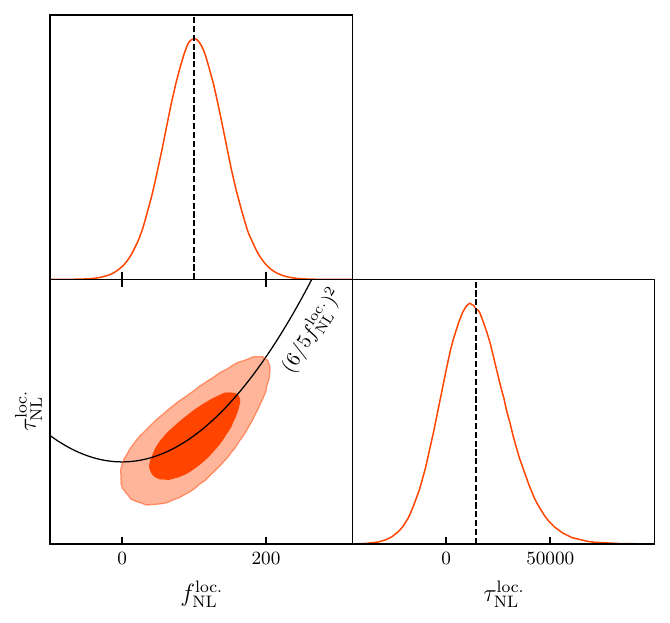}
\caption{Constraints on $\tau_{\rm NL}^{\rm loc.}$ and $f_{\rm NL}^{\rm loc.}$ from a joint analysis of the squeezed matter bispectrum and collapsed matter trispectrum measured from the \textsc{Quijote}-PNG simulations, with $f_{\rm NL}^{\rm loc.}=100$. The black line shows the relationship between $\tau_{\rm NL}^{\rm loc.}$ and $f_{\rm NL}^{\rm loc.}$ for single-source local non-Gaussianity. Using the collapsed trispectrum, we can jointly infer $f_{\rm NL}^{\rm loc.}$ and $\tau_{\rm NL}^{\rm loc.}$. These results use the $z=0$ measurements with the covariance estimated for a $(1~{\rm Gpc}/h)^3$ box.} \label{fig:tau_NL}
\end{figure}

We now discuss the information from the collapsed trispectrum. In Fig.~\ref{fig:trispectrum_SV_cancellation}, we show the error on $f_{\rm NL}^\Delta$ from a joint analysis of the squeezed matter bispectrum and the collapsed matter trispectrum. We analyze the bispectrum and trispectrum measurements binned in soft modes with $0.005<q<0.05~h/{\rm Mpc}$, and split in two equal-width hard mode bins with $k_{\rm min}=60k_f\approx 0.38~h/{\rm Mpc}$ and $k_{\rm max}=320k_f\approx 2.01~h/{\rm Mpc}.$ We only show results for the $\Delta=0.5$ simulations, but we find similar results for all simulations considered in this work. The solid lines show the results from our fiducial analysis where we assume a joint likelihood in the measured soft matter power spectrum to cancel cosmic variance. For such an analysis, the constraint on $f_{\rm NL}^\Delta$ improves negligibly after including the collapsed trispectrum. As discussed in Ref.~\cite{dePutter:2018jqk}, analyzing the bispectrum in multiple $k$-bins jointly with the measured matter power spectrum leads to near-perfect cosmic variance cancellation. Thus, the collapsed matter trispectrum provides little information on $f_{\rm NL}^\Delta$ beyond the squeezed matter bispectrum and soft matter power spectrum. To further illustrate the importance of sample variance cancellation in the matter sector, we repeat this analysis assuming no knowledge of the particular realization of $P_m(q).$ In this case, including the collapsed trispectrum (red dashed) leads to considerably tighter constraints on $f_{\rm NL}^\Delta$ than an analysis of the squeezed bispectrum alone (black dashed).

Finally, even if the collapsed trispectrum does not significantly improve the constraint on $f_{\rm NL}^\Delta$ (if knowledge of the power spectrum is available), it can be used to constrain more general models of PNG where multiple fields contribute to the curvature perturbation. As an example of this, Fig.~\ref{fig:tau_NL} shows the marginalized constraints on $f_{\rm NL}^{\rm loc.}$ and $\tau_{\rm NL}^{\rm loc.}$ derived from a joint analysis of measurements of the squeezed matter bispectrum and collapsed trispectrum.  This fit uses the squeezed matter bispectrum and collapsed matter trispectrum measured in two equal-width $k$-bins with $ 0.37<k<1.88~h/{\rm Mpc}$. We model the collapsed trispectrum including the $\tau_{\rm NL}^{\rm loc.}$ contribution using Eq.~\eqref{eq:tau_NL_collapsed_T}. We use the mean measurements from 50 realizations of the \textsc{Quijote} simulations with $f_{\rm NL}^{\rm loc.}=100$ and the covariance estimated for a $(1~{\rm Gpc}/h)^3$ volume. The solid black line shows the relationship between $f_{\rm NL}^{\rm loc.}$ and $\tau_{\rm NL}^{\rm loc.}$ in quadratic local non-Gaussianity. We infer the true value of $f_{\rm NL}^{\rm loc.}$ and $\tau_{\rm NL}^{\rm loc.}.$ Thus, the collapsed trispectrum provides a powerful probe of higher-order PNG.

\subsection{Scale-dependent halo bias}\label{subsec:res_scale_dep_bias}

\begin{figure}[!t]
\centering
\includegraphics[width=0.99\linewidth]{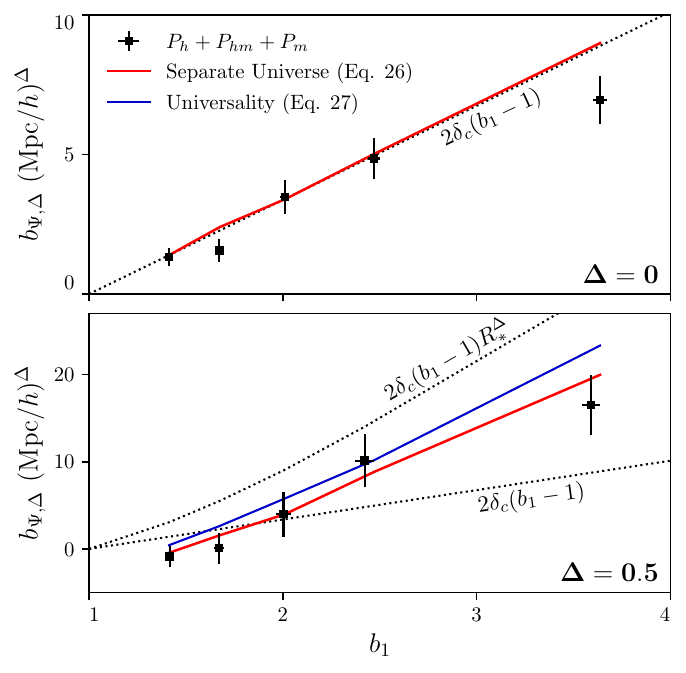}
\caption{Constraints on the $b_1-b_{\Psi,\Delta}$ relationship from PNG simulations with $\Delta=0$ (top) and $\Delta=0.5$ (bottom). The black data points are derived by fitting $P_{h}$, $P_{hm}$, and $P_{m}$ jointly in five different halo mass bins with $f_{\rm NL}^\Delta$ and $\Delta$ fixed at their true values. The solid red line, which is computed from separate universe simulations using Eq.~\eqref{eq:b_psi_SU}, is in excellent agreement with the measured bias. The measured bias is also consistent with the universality prediction (Eq.~\ref{eq:non_Gaussian_bias_UMF}), although this prediction is likely inaccurate for more realistic halo and galaxy samples~\cite{Barreira:2020kvh, Barreira:2021ueb, Lazeyras:2022koc,Barreira:2022sey, Barreira:2023rxn}. At large values of $b_1$, $b_{\Psi}^{\Delta=0.5}$ is significantly larger than $b_{\Psi}^{\Delta=0}$, illustrating the impact of the Cosmological Collider bispectrum on galaxy formation.} \label{fig:b1_bphi_relation}
\end{figure}

\noindent We now analyze the scale-dependent halo bias of our simulations. Fig.~\ref{fig:b1_bphi_relation} shows the $b_1-b_{\Psi,\Delta}$ relationship for mass-selected halos from the $\Delta=0$ (top) and $\Delta=0.5$ (bottom) simulations at the redshift $z=0$.\footnote{We do not analyze the scale-dependent bias in the $\Delta=1$ simulations because of the large error on the values of $b_{\Psi,\Delta}$ fit to these simulations.} The black points denote the mean values of $b_1$ and $b_{\Psi,\Delta}$, alongside their 68\% confidence intervals, inferred from a joint analysis of the large-scale halo auto-power spectrum ($P_h$(q)), halo-matter cross-power spectrum $(P_{hm}(q))$, and matter power spectrum $P_{m}(q)$ averaged over 50 realizations. For this analysis, we fix $f_{\rm NL}^\Delta$ and $\Delta$ to their true values for a given simulation. Finally, since we model the halo statistics using linear theory, we only include modes with $0.006<q<0.025~h/{\rm Mpc}$.

The top panel of Fig.~\ref{fig:b1_bphi_relation} illustrates the $b_1-b_{\Psi,\Delta}$ relationship for local PNG ($\Delta=0$). We find excellent agreement between the measured non-Gaussian bias parameters and the predictions from the separate universe simulations (solid red line) and universality (dotted black line). Note that for $\Delta=0$, the universality prediction (Eq.~\ref{eq:non_Gaussian_bias_UMF}) is equivalent to $b_{\Psi,\Delta}=2\delta_c(b_1-1)$. The bottom panel shows the same results for the $\Delta=0.5$ simulations. As in the $\Delta=0$ simulations, the measured non-Gaussian bias is consistent with the theoretical predictions from the separate universe simulations (solid red) and universality (solid blue). Notably, for non-zero values of $\Delta$, the universality prediction for $b_{\Psi,\Delta}$ depends on the Lagrangian radius of the halo, $R_*(M)$. In this work, we estimate $R_*$ from the mean FoF mass of each halo sample. Although we find that $b_{\Psi,\Delta}$ agrees with the universality prediction for the halos considered here, we emphasize that the main takeaway from Fig.~\ref{fig:b1_bphi_relation} is the agreement between the separate universe predictions and measured bias parameters. For more realistic halo and galaxy samples, the universality prediction is likely a poor approximation of $b_{\Psi,\Delta}$~\cite{Barreira:2020kvh, Barreira:2021ueb, Lazeyras:2022koc, Barreira:2022sey, Barreira:2023rxn}.

In the bottom panel of Fig.~\ref{fig:b1_bphi_relation}, the lower dashed black line shows the universality prediction for local PNG ($\Delta=0$). For large values of $b_1$, the non-Gaussian bias for the $\Delta=0.5$ cosmology is significantly enhanced relative to that of the $\Delta=0$ cosmology, akin to the non-linear enhancement of the potential derivatives seen in Fig.~\ref{fig:response_function}. The upper dashed black line in the same panel is a simple prediction for the non-Gaussian bias based on dimensional analysis, where $b_{\Psi,\Delta}=2\delta_c(b_1-1) R_*^\Delta$. This prediction overestimates the non-Gaussian bias for the halos considered here. Nevertheless, it captures the approximate shape of the $b_1-b_{\Psi,\Delta}$ relationship for this halo sample.

Stepping back, these findings demonstrate the utility of leveraging simulations with analytic calculations to study the influence of intermediate-mass particles during inflation on the large-scale halo bias. Using separate universe simulations, we can accurately predict the non-Gaussian bias parameters for specific halo samples. These results have several potential applications, particularly in the context of generalizing studies of $b_\phi$ in local PNG, to studies of $b_{\Psi,\Delta}$ for the models considered here. Given the significant variation in the $b_{\Psi,\Delta}(b_1)$ relationship for the $\Delta=0.5$ simulations, one could carefully choose galaxy samples to optimally constrain $f_{\rm NL}^\Delta.$ Additionally, exploring higher-order statistics of the halo field by generalizing the methods we developed for the matter field could be interesting. We choose not to study the information content of the halo field in this work because previous work has found that, for the number densities used in the \textsc{Quijote} simulations, the squeezed halo bispectrum provides little information beyond the scale-dependent halo bias~\cite{Coulton:2022rir, Giri:2022nzt}. Nevertheless, comparing the constraining power of the scale-dependent halo bias with other summary statistics for the models considered here would be an interesting follow-up.

\section{Conclusion}\label{Sec:conclusions}

\noindent Fields with mass $m\approx H$ present during inflation produce distinct signatures in the squeezed bispectrum of the primordial curvature perturbation~\cite{Arkani-Hamed:2015bza,Assassi:2012zq,Chen:2009zp,Lee:2016vti}. Searching for these shapes of PNG presents a unique opportunity to uncover the spectrum of particles present during inflation. In this work, we have presented non-perturbative techniques to search for light and intermediate-mass fields ($0\leq m<3H/2$) present during inflation using LSS data. Our main results are summarized as follows:
\begin{itemize}
    \item  We run $N$-body simulations with PNG sourced by intermediate-mass scalars during inflation, using the same settings as \textsc{Quijote-PNG}~\cite{Villaescusa-Navarro:2019bje, Coulton:2022rir}, thus complementing existing $N$-body simulations with PNG. These simulations could be a valuable tool for future studies of the Cosmological Collider.
    \item Using the separate universe approach, we derive non-perturbative models for the squeezed matter bispectrum (Eq.~\ref{eq:squeezed_B_non_pert_collider_model}) and collapsed matter trispectrum (Eq.~\ref{eq:trispectrum_no_tauNL}) associated with this type of PNG. We use these models to estimate the amplitude of PNG, $f_{\rm NL}^\Delta$, from $N$-body simulations deep into the non-linear regime ($k_{\rm max}=2~h/{\rm Mpc}$ at $z=0$).
    \item For light fields ($\Delta=0$), the constraint on $f_{\rm NL}^\Delta$ improves by $\approx 40\%$ after increasing $k_{\rm max}$ from $1$ to $2~h/{\rm Mpc}$. The improvement is smaller for heavier fields $(\Delta>0$) because the bispectrum is suppressed by $\approx k^{-\Delta}$. Additionally, including the collapsed trispectrum provides a negligible improvement in our constraint on $f_{\rm NL}^\Delta$ (due to sample-variance cancellation in the bispectrum), but allows us to constrain higher-order PNG, such as $\tau_{\rm NL}^{\rm loc.}$.
    \item The error on $f_{\rm NL}^\Delta$ grows significantly as $\Delta$ increases. This is due to both the milder scaling of the pole, as well as the increased degeneracy between $f_{\rm NL}^\Delta$ and the nuisance parameters in our model for $\Delta>0.$ 
    \item We study the scale-dependent halo-bias sourced by intermediate-mass particles during inflation. The measured non-Gaussian bias parameter agrees with theoretical predictions from (i) separate universe simulations and (ii) assuming universal mass functions, although the second prediction will likely fail for realistic galaxy samples.
\end{itemize}

There are several interesting follow-ups to this analysis. On the simulation side, it would be interesting to extend our formalism to spinning fields, which lead to an orientation-dependent bispectrum, as well as fields with $m>3H/2$, which induce oscillations in the squeezed bispectrum. These squeezed bispectrum configurations could be simulated by generalizing the coupling kernel in Eq.~\eqref{eq:kernel_scale_inv}. We discuss this in more detail in Appendix~\ref{sec:App_spin_and_osc}. Previous works have presented methods to search for spinning fields using galaxy shape statistics~\cite{Schmidt:2015xka, Akitsu:2020jvx}. It would be interesting to quantify how competitive these are with constraints from the squeezed bispectrum and collapsed trispectrum. Similarly, the methods presented here could be used to search for tachyons during inflation, which produce an even stronger scaling in the squeezed bispectrum than local PNG~\cite{McCulloch:2024hiz}. 

On the observational front, we could apply these methods to the (CMB or galaxy) weak lensing bispectrum, although our previous work has shown that this does not provide competitive constraints on $f_{\rm NL}^{\rm loc.}$ for realistic upcoming surveys~\cite{Goldstein:2023brb}. The situation is expected be even worse for large values of $\Delta$ due to the severe penalty from marginalizing over nuisance parameters. Thus, it would be useful to have informative priors on the nuisance parameters in this model. For the non-linear matter statistics considered here, the main issue arises from marginalizing over $\bar{a}_{0}^{(i)}$. In principle, this can be computed from separate universe simulations~\cite{Chiang:2017jnm, Biagetti:2022ckz}, or approximated from theory~\cite{Valageas:2016hhr}, although more work is necessary to assess the robustness of these methods. It is also possible that one can break these degeneracies by including non-squeezed bispectrum configurations, although the methods presented here only work in the squeezed limit, and the non-squeezed signatures are more easily mimicked by single-field effects. 

Given that this work focused primarily on constraining $f_{\rm NL}^\Delta$ using the squeezed matter bispectrum and collapsed matter trispectrum, it would be interesting to extend our models for the matter field statistics to describe the real-space and redshift-space squeezed galaxy bispectrum and collapsed galaxy trispectrum. In particular, it would be useful to quantify the information content of each of these statistics, as it remains to be seen whether higher-order statistics of the halo field provide any significant improvement over constraints on $f_{\rm NL}^\Delta$ from the scale-dependent bias~\cite{Giri:2022nzt}. Alternatively, given the challenges posed by large-angle systematics in galaxy surveys, a potentially interesting follow-up would be to use our models to derive non-perturbative quadratic-estimator approaches for constraining PNG, which can reconstruct the long-wavelength modes from small-scale correlations~\cite{Darwish:2020prn, Wang:2023lvt}.

There are also several interesting follow-ups investigating the impact of the Cosmological Collider on small-scale structure. For example, it could be useful to have a more physical understanding of the behavior of the non-linear response functions and non-Gaussian bias parameters estimated from our separate universe simulations.\footnote{It could be interesting to study the impact of the Cosmological Collider bispectrum on the halo mass function and halo density profiles. This would require higher-resolution simulations than those used here, as well as a more careful procedure to generate the initial conditions to account for non-squeezed bispectrum configurations. Such a study would complement the recent findings of Ref.~\cite{Baldi:2024okt}, which used simulations to analyze the impact of scale-dependent PNG on non-linear structure. Although we note that the squeezed limit of the bispectrum used in Ref.~\cite{Baldi:2024okt} is enhanced by the hard mode $k$ (and independent of the soft mode $q$), whereas the template considered here is suppressed by the ratio $(q/k)^\Delta$.} A good understanding of $b_{\Psi,\Delta}$ from, \emph{e.g.}, hydrodynamical simulations, could significantly improve our ability to constrain these models, especially in multi-tracer scenarios~\cite{Sullivan:2023qjr}. On a similar note, the simulations developed here could be used to test more formal aspects of the non-Gaussian bias parameters, such as their running~\cite{Nikolis:2024kbx}. 

Finally, the ultimate dream of Cosmological Collider Physics is not to constrain the amplitude, $f_{\rm NL}^\Delta$, but rather to determine the shape of the pole, 
$\Delta$, and consequently the particle mass. Ideally, this should be done as a joint analysis with other shapes of $f_{\rm NL}$, which parameterize the background of inflationary particles. Unfortunately, as there is currently no detection of PNG, there is no background to marginalize over. Thus, just like in true collider physics, we are left anxiously waiting.
\hspace{10pt}
\vspace{15pt}

\acknowledgments

\noindent We are grateful to Fabian Schmidt for insightful discussions and to Paco Villaescusa-Navarro for help running the \textsc{Quijote} simulations. OHEP is a Junior Fellow of the Simons Society of Fellows and acknowledges Aperol for providing creative inspiration.  JCH acknowledges support from NSF grant AST-2108536, NASA grant 80NSSC22K0721, NASA grant 80NSSC23K0463, DOE grant DE-SC0011941, the Sloan Foundation, and the Simons Foundation. LH acknowledges support from the DOE DE-SC011941 and a Simons Fellowship in Theoretical Physics.
The authors acknowledge the Texas Advanced Computing Center (TACC)\footnote{\href{http://www.tacc.utexas.edu}{http://www.tacc.utexas.edu}} at The University of Texas at Austin for providing computational resources that have contributed to the research results reported within this paper. We acknowledge computing resources from Columbia University's Shared Research Computing Facility project, which is supported by NIH Research Facility Improvement Grant 1G20RR030893-01, and associated funds from the New York State Empire State Development, Division of Science Technology and Innovation (NYSTAR) Contract C090171, both awarded April 15, 2010.

\section*{Data Availability}

\noindent The simulations used in this work are publicly available. More details can be found at \href{https://quijote-simulations.readthedocs.io/en/latest/collider.html}{https://quijote-simulations.readthedocs.io/en/latest/collider.html}.

\bibliographystyle{apsrev4-1}
\bibliography{biblio}

\onecolumngrid
\clearpage
\appendix

\section{Cosmological Collider initial conditions including spin and massive-er particles}\label{sec:App_spin_and_osc}

\noindent In this appendix, we describe how to generate $N$-body initial conditions with the Cosmological Collider squeezed bispectrum template for fields with spin, as well as masses $m>3/2H$. Importantly, the transformations derived here can be efficiently computed using Fast Fourier Transformations (FFTs).

\subsection{Intermediate-mass spinning particles}
\noindent For intermediate-mass fields with even spin, $s$,\footnote{The squeezed bispectrum suppressed by an additional power of $(q/k)$ for odd values of $s$~\cite{MoradinezhadDizgah:2017szk, MoradinezhadDizgah:2018ssw} and thus ignored herein.} the squeezed bispectrum depends on the angle between the momenta $\vq$ and $\vk$,
\begin{equation}\label{eq:squeezed_bk_qsfi_spin}
\lim_{q\ll k}B_{\Phi}(\vq,\vk) = 4 f_{\rm NL}^{\Delta}\left(\frac{q}{k}\right)^\Delta \mathcal{L}_s(\hat{\vq}\cdot\hat{\vk})P_{\Phi}(q)P_{\Phi}(k),
\end{equation}
where $\mathcal{L}_s(x)$ is the Legendre polynomial of order $s$ and the (real) exponent is,\footnote{Unless $\Delta=3/2$, the trispectrum in such models takes a more complex form than that of Eq.~\eqref{eq:collapsed_T_no_contact} \citep{Arkani-Hamed:2015bza}, due to the massive field polarization states. Treatment of this is beyond the scope of this paper.} 
\begin{equation}
        \Delta=
        \begin{cases} 3/2-\sqrt{9/4-m^2/H^2} &\textrm{ for } s=0, \\
        3/2-\sqrt{\left(s-{1}/{2}\right)^2-m^2/H^2} &\textrm{ for }s\neq 0.
        \end{cases}
\end{equation}
Following Sec.~\ref{subsec:PNG_IC_sims}, we can simulate this squeezed bispectrum using the kernel
\begin{align}\label{eq:kernel_with_spin}
    \begin{split}
K_{\Delta}(\vk_1, \vk_2)&=\left[\left(\frac{k_1}{k_{12}}\right)^{\Delta}+\left(\frac{k_2}{k_{12}}\right)^{\Delta}-1\right]\mathcal{L}_s(\hat{\vk_1}\cdot \hat{\vk_2})\,, \\
        &=\left(\frac{4\pi}{s+1}\right)\sum_{m=-s}^{s}\left[\left(\frac{k_1}{k_{12}}\right)^{\Delta}+\left(\frac{k_2}{k_{12}}\right)^{\Delta}-1\right]Y_{sm}(\hat{\vk_1})\,Y_{sm}^*(\hat{\vk_2})\,,
    \end{split}
\end{align}
where we separated the kernel using the spherical harmonic addition theorem.\footnote{For a fixed value of $s$, one could also derive a coupling kernel by writing out the Legendre polynomial of order $s$ explicitly instead of using spherical harmonics. Ref.~\cite{Akitsu:2020jvx} does this for massless spin-2 fields.} 
The auxiliary field in Eq.~\eqref{eq:quadratic_png} can be efficiently evaluated using FFTs
\begin{equation}
    \Psi_\Delta(\vk)=\frac{4\pi}{2s+1}\int d^3x\,e^{i\vk\cdot\vx}\sum\limits_{m=-s}^s(-1)^m\left[ \frac{2}{k^\Delta}\mathcal{F}_{s,m}^{\,\Delta}(\bx)\mathcal{F}_{s,-m}^{\,0}(\bx)-\mathcal{F}_{s,m}^{\,0}(\bx)\mathcal{F}_{s,-m}^{\,0}(\bx)\right],
\end{equation}
where 
\begin{equation}
    \mathcal{F}_{s,m}^{\,\gamma}(\bx)\equiv \int_\vp p^\gamma\,\phi(\vp)Y_{sm}(\hat{\vp})\, e^{-i\vp \cdot \vx}.
\end{equation}

The squeezed bispectrum associated with Eq.~\eqref{eq:kernel_with_spin} is
\begin{align}
    \lim_{q\ll k}B_\Phi(\vq, \vk)&=4f_{\rm NL}^\Delta\left[\left(\frac{q}{k} \right)^\Delta{\mathcal{L}_s(\hat{\vq}\cdot \hat{\vk})}P_\Phi(q)P_\Phi(k)+\left(\frac{k}{q}\right)^\Delta P_\Phi(k)P_\Phi(k)\right],\\
    &\approx 4f_{\rm NL}^\Delta\left[\left(\frac{q}{k} \right)^\Delta{\mathcal{L}_s(\hat{\vq}\cdot \hat{\vk})}+\left(\frac{q}{k}\right)^{3-\Delta} \right]P_\Phi(q)P_\Phi(k),
\end{align}
where the second line assumes $P_{\Phi}(k)$ is scale-invariant. Thus, the kernel generates the target squeezed bispectrum if $\Delta \ll 3/2.$ On the other hand, particles with $s\neq 0$ must satisfy the Higuchi bound, $m^2/H^2\geq s(s-1)$~\cite{Higuchi:1986py}, which implies $\Delta\geq 1$. Therefore, the second term in the squeezed bispectrum non-negligible for the majority of the physical parameter space for spin-$s$ particles. One can suppress this unwanted contribution by using a high-pass filtered kernel $K_{\Delta}(\vk_1,\vk_2)\rightarrow K_\Delta(\vk_1,\vk_2)\Theta(k_{12}-k_L)$, where $\Theta$ is the step function and $k_L$ is above all $q$-values of interest (e.g., $k_L\approx 0.1~h/{\rm Mpc}$ in this work). Alternatively, if $\Delta=3/2$, one can get rid of the unwanted contribution by subtracting off the kernel for intermediate-mass scalars with $\Delta=3/2$, \emph{i.e.}, subtracting Eq.~\eqref{eq:kernel_scale_inv} with an additional factor of $1/2$.

\subsection{Oscillatory bispectra from massive scalars}

\noindent For particles with mass $m>3H/2$, the squeezed bispectrum oscillates logarithmically as
\begin{equation}\label{eq:squeezed_bk_qsfi_spin}
\lim_{q\ll k}B_{\Phi}(\vq,\vk) = 4 f_{\rm NL}^{\Delta}\left(\frac{q}{k}\right)^{3/2}\cos\left(\nu\log\left(\frac{q}{k}\right)+\delta\right)\mathcal{L}_s(\hat{\vq}\cdot\hat{\vk})P_{\Phi}(q)P_{\Phi}(k),
\end{equation}
where $\nu=\sqrt{m^2/H^2-9/4}$ for scalars and $\nu=\sqrt{m^2/H^2-(s-1/2)^2}$ for $s\neq 0.$ The phase ($\delta$) is a function of the mass and spin, and depends on the precise inflationary coupling (see~\cite{Lee:2016vti,Sohn:2024xzd}). For simplicity, we focus on scalar fields, but note that these results can be generalized to spinning fields using the procedure developed for intermediate-mass spinning particles. For scalar fields, we consider the following kernel:
\begin{align}\label{eq:massiver_scalar_kernel}
\begin{split}
    K_\Delta(\vk_1,\vk_2)&=\left\{\left(\frac{k_1}{k_{12}} \right)^{3/2} \cos\left(\nu\log\left(\frac{k_1}{k_{12}}\right)+\delta\right)+\left(\frac{k_2}{k_{12}} \right)^{3/2} \cos\left(\nu\log\left(\frac{k_2}{k_{12}}\right)+\delta\right)-\cos(\delta) \right\},\\
    &=\frac{1}{2}\left\{ e^{i\delta}\left[\left(\frac{k_1}{k_{12}}\right)^{3/2+i\nu}+\left(\frac{k_2}{k_{12}}\right)^{3/2+i\nu}-1 \right]+e^{-i\delta}\left[\left(\frac{k_1}{k_{12}}\right)^{3/2-i\nu}+\left(\frac{k_2}{k_{12}}\right)^{3/2-i\nu}-1 \right]\right\},
\end{split}
\end{align}
where the second line is written in a separable form that can be easily evaluated using FFTs. For scale-invariant $P_{\Phi}(k)$, the squeezed bispectrum associated with Eq.~\eqref{eq:massiver_scalar_kernel} is, up to $\mathcal{O}\left(f_{\rm NL}^\Delta \right)$,
\begin{align}\label{eq:scalar_osc_bispectrum}
\begin{split}
\lim_{q\ll k}B_{\Phi}(\vq,\vk) =&4 f_{\rm NL}^{\Delta}\left(\frac{q}{k}\right)^{3/2}\left[\cos\left(\nu\log\left(\frac{q}{k}\right)+\delta\right) +\cos\left(\nu\log\left(\frac{q}{k}\right)-\delta\right)\right]P_{\Phi}(q)P_{\Phi}(k).
\end{split}
\end{align}
Similar to the intermediate-mass case, we can apply a high-pass filter to the kernel to suppress the second term in Eq.~\eqref{eq:scalar_osc_bispectrum}. Alternatively, one could always fix $\delta=0$ and multiply the kernel by $1/2$. Although this $\delta=0$ template lacks physical motivation, simulations with $\delta=0$ would likely suffice to study the impact of oscillatory bispectra on large-scale structure in the non-linear regime.

\vfill

\pagebreak

\section{Bispectrum and trispectrum estimators}\label{sec:App_estimators}

\noindent In this appendix, we derive estimators for the 3D real-space bispectrum and trispectrum from a simulation volume. These estimators can be efficiently computed by Fast Fourier Transformations (FFTs). We implement these estimators in the publicly available code \texttt{PNGolin}.\footnote{\href{https://github.com/samgolds/PNGolin}{https://github.com/samgolds/PNGolin}}

\subsection{Bispectrum estimator}

\noindent The bispectrum is parameterized by three external momenta, $k_1$, $k_2$, and $k_3$. To estimate the bispectrum from a density field, $\delta_\vk\equiv \delta(\vk)$, we introduce a window $W_{k_i}(p)$ which is equal to $1$ if $p$ is in $k_i$ and zero otherwise. A bispectrum estimator is then
\begin{align}\label{eq:bispectrum_estimator}
    \hat{B}(k_1, k_2, k_3)\propto \int\limits_{\vp_1,\dots, \vp_3}\bigg[&W_{k_1}(p_1)W_{k_2}(p_2)W_{k_3}(p)\,\delta_{\vp_1}\delta_{\vp_2}\delta_{\vp_3}(2\pi)^3\delta_D^{(3)}(\vp_{123})\bigg],
\end{align}
where we have ignored the overall normalization, which can be computed by re-evaluating the integral with all density fields replaced by the identity. To proceed, we write the Dirac delta as an exponential integral,

\begin{equation}
    \delta_D(\vk)=\frac{1}{(2\pi)^3}\int d^3x e^{-i\vk\cdot \bx}.
\end{equation} 
The bispectrum estimator is then
\begin{equation}\label{eq:B_est_sep}
\hat{B}(k_1,k_2,k_3)\propto\int d^3x\left[\left(\int_{\vp_1} e^{-i\vp_1\cdot \vx}\,W_{k_1}(p_1)\,\delta_{\vp_1} \right)\left(\int_{\vp_2} e^{-i\vp_2\cdot \vx}\,W_{k_2}(p_2)\,\delta_{\vp_2} \right)\left(\int_{\vp_3} e^{-i\vp_3\cdot \vx}\,W_{k_3}(p_3)\,\delta_{\vp_3} \right)\right],
\end{equation}
which can be efficiently evaluated using FFTs. In this work, we analyze the squeezed bispectrum with $k_2=k_3$, making the last two inverse Fourier transforms in Eq.~\eqref{eq:B_est_sep} equal.

\subsection{Trispectrum estimator}

\noindent Our trispectrum estimator is based on the parity-even trispectrum estimator described in Appendix A of Ref.~\cite{Coulton:2023oug}. The trispectrum can be parameterized by four external momenta, $k_1$, $k_2$, $k_3$, and $k_4$, as well as the two internal momenta, $K\equiv k_{12}$ and $K'=k_{14}$ (see Fig.~\ref{fig:squeezed_B_collapsed_T_shape_dependence}). The full trispectrum estimator is not separable, thus we integrate over $k_{14}$ and estimate the trispectrum with external momenta in bins $k_1,\dots k_4$ and internal momentum $k_{12}\in K$.\footnote{See Ref.~\cite{Gualdi:2020eag} for an estimator that is also integrated over $k_{12}.$} An estimator for the total four-point function, \emph{i.e.}, including disconnected contributions, is
\begin{align}\label{eq:total_estimator_Tk}
    \hat{T}_{\rm tot.}(k_1, k_2, k_3, k_4, K)\propto \int\limits_{\vP}W_K(P)\int\limits_{\vp_1,\dots, \vp_4}\bigg[&W_{k_1}(p_1)W_{k_2}(p_2)W_{k_3}(p_3)W_{k_4}(p_4)\,\delta_{\vp_1}\delta_{\vp_2}\delta_{\vp_3}\delta_{\vp_4}\\
    & (2\pi)^3\delta_D^{(3)}(\vp_{12}-\vP)(2\pi)^3\delta_D^{(3)}(\vp_{34}+\vP)\bigg]\nonumber,
\end{align}
where we have again ignored the overall normalization, which can be computed by re-evaluating the integral with all density fields replaced by the identity. Eq.~\eqref{eq:total_estimator_Tk} can be efficiently evaluated using FFTs by replacing the Dirac delta functions with exponential integrals, yielding
\begin{equation}\label{eq:full_4pcf_estimator}
  \hat{T}_{\rm tot.}(k_1, k_2, k_3, k_4, K)\propto \int_{\vP}W_K(P) \left( \int d^3x\,e^{-i\vP\cdot \bx} \delta_{W_{k_1}}(\bx) \delta_{W_{k_2}}(\bx) \right) \left( \int d^3y\,e^{i\vP\cdot \by} \delta_{W_{k_3}}(\by) \delta_{W_{k_4}}(\by) \right),
\end{equation}
where $ \delta_{W_{k_i}}(\bx)\equiv \int_{\vp}W_{k_i}(p)\delta_{\vp}e^{i\vp\cdot\bx}$.

\pagebreak

\begin{figure*}[!t]
\centering
\includegraphics[width=0.995\linewidth]{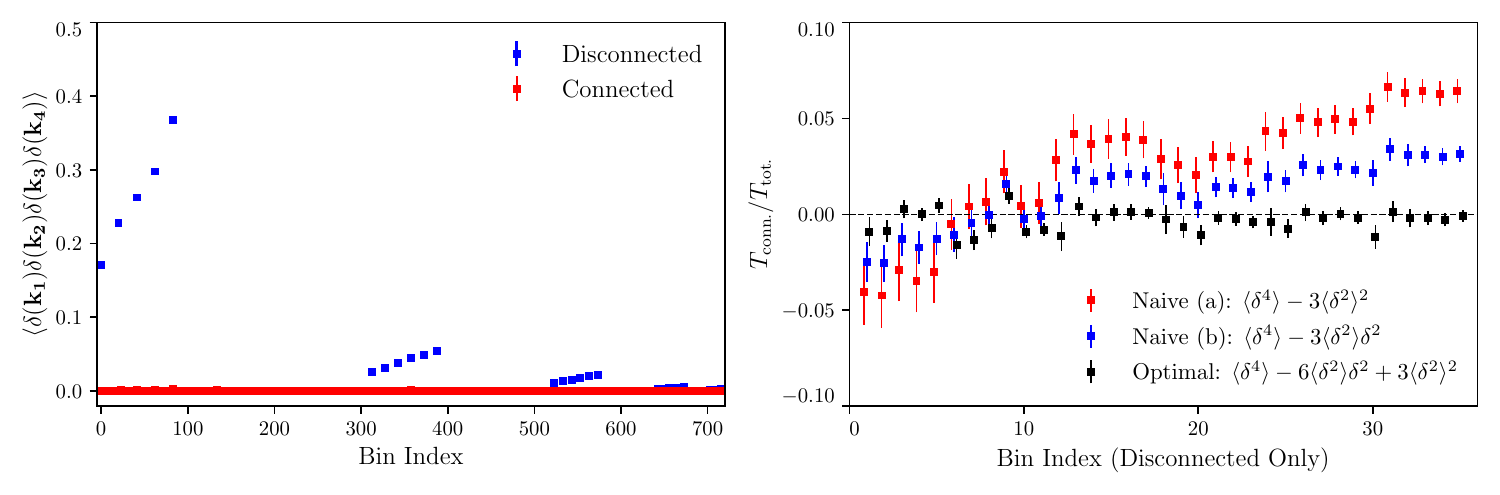}
\caption{Validation of trispectrum estimator on 200 simulated Gaussian random fields with a scale-invariant power spectrum. The left panel shows the connected (red) and disconnected (blue) contributions to the four-point function estimator (Eq.~\eqref{eq:total_estimator_Tk}). The disconnected contributions are large and must be subtracted to estimate the trispectrum. The trispectrum (red) is consistent with zero. The right panel shows the ratio of the trispectrum ($T_{\rm conn.}$) to the total four-point function ($T_{\rm tot.})$ for several estimates of the disconnected term. The optimal estimator (black) has lower bias and lower variance than the naive estimators (blue and red). Note that since we generate the Gaussian random fields with a \emph{known} power spectrum, the bias is due to finite grid size effects. We discuss this in more detail in the main text.}
\label{fig:trispectrum_disconnected_test}
\end{figure*}

 Eq.~\eqref{eq:total_estimator_Tk} includes disconnected contributions that must be subtracted to estimate the trispectrum. In the limit of mild non-Gaussianity, the optimal estimator for the trispectrum is~\cite{Smith:2015uia, Shen:2024vft},
 \begin{equation}
     \hat{T}_{\rm opt}=\delta^4-6\,\delta^2\langle \delta^2\rangle+3\langle \delta^2\rangle^2.
 \end{equation}
 
 Therefore, we can derive an optimal estimator for the trispectrum by replacing $\delta^4 \rightarrow \delta^4-6\,\delta^2\langle \delta^2\rangle+3\langle \delta^2\rangle^2$ in Eq.~\eqref{eq:total_estimator_Tk}. The $3\langle \delta^2\rangle^2$ term corresponds to a \emph{realization-independent} disconnected contribution, which is given by
\begin{equation}\label{eq:disc_real_indep}
    \hat{T}_{\rm disc.}^{\,3\langle \delta^2 \rangle^2} \propto \int\limits_{\vP}W_K(P)\left( \prod\limits_{i=1}^4\int\limits_{\vp_i}W_{k_i}(p_i)\right) \left[ P(p_1)P(p_3)\,\delta_D^{(3)}(\vp_{12})\,\delta_D^{(3)}(\vp_{34})+2~{\rm perm.}\right]
     \delta_D^{(3)}(\vp_{12}-\vP)\delta_D^{(3)}(\vp_{34}+\vP),
\end{equation}
where $P(p_i)$ is calculated from a fiducial power spectrum and we have omitted the momentum dependence on the left-hand side for clarity. Assuming $\vK\neq 0$, the first term in Eq.~\eqref{eq:disc_real_indep} is zero. The remaining two permutations are
\begin{align*}
 \hat{T}_{\rm disc.}^{\,3\langle \delta^2 \rangle^2}&\propto \int\limits_{\vP}W_K(P)\left( \prod\limits_{i=1}^4\int\limits_{\vp_i}W_{k_i}(p_i)\right) \left[ P(p_1)P(p_2)\,\delta_D^{(3)}(\vp_{13})\,\delta_D^{(3)}(\vp_{24})\right]
     \delta_D^{(3)}(\vp_{12}-\vP)\delta_D^{(3)}(\vp_{34}+\vP)+(3\leftrightarrow 4), \\
     & \propto \int_{\vP}W_K(P) \left( \int_{\vp_1}W_{k_1}(p_1)W_{k_3}(p_1)P(p_1)\right) \left( \int_{\vp_2}W_{k_2}(p_2)W_{k_4}(p_2)P(p_2)\right)\left[(2\pi)^3\delta_D^{(3)}(\vp_{12}-\vP)\right]^2+(3\leftrightarrow 4).
\end{align*}
To proceed, we approximate the squared Delta function using $((2\pi)^3\delta^{(3)}_D(\vk))^2=V\times (2\pi)^3\delta^{(3)}_D(\vk)$ (see, \emph{e.g.},~\cite{Gualdi:2020eag}) and rewrite the final Delta function with an exponential integral, yielding
\begin{equation}\label{eq:T_disc_3_PP}
\hat{T}_{\rm disc.}^{\,3\langle \delta^2\rangle^2}=\frac{V}{ N_{k_1,k_2,k_3,k_4,K}}\int_{\vP}W_K(P)  \int d^3x\, e^{-i\vP\cdot \bx}\left[ F^{P}_{13}(\bx)F^{P}_{24}(\bx)+F^{P}_{14}(\bx)F^{P}_{23}(\bx)\right], 
\end{equation}
where
\begin{equation}
    F^{P}_{ij}(\bx)\equiv \int_\vp W_{k_i}(p)W_{k_j}(p)P(p)e^{i\vp\cdot\bm{x}},
\end{equation}
and $N_{k_1,k_2,k_3,k_4,K}$ is the number of tetrahedra in a bin. 

Following the same reasoning, the $6\delta^2\langle \delta^2\rangle$ term corresponds to a \emph{realization-dependent} disconnected contribution,
\begin{align}\label{eq:T_disc_6_deltaP}
    \hat{T}_{\rm disc.}^{6\,\delta^2\langle \delta^2\rangle}=\frac{V}{ N_{k_1,k_2,k_3,k_4,K}}\int_{\vQ}W_E(Q)  \int d^3x\, e^{-i\vQ\cdot \bx}[& F^P_{13}(\bx)F^\delta_{24}(\bx)+F^\delta_{13}(\bx)F^P_{24}(\bx)\notag\\
    &+F^P_{14}(\bx)F^\delta_{23}(\bx)+F^\delta_{14}(\bx)F^P_{23}(\bx)],
\end{align}
where
\begin{equation}
    F^\delta_{ij}(\bx)\equiv \int_\vk W_{k_i}(p)W_{k_j}(p)|\delta_{\vp}|^2e^{i\vk\cdot\bm{x}}.
\end{equation}
To summarize, an optimal estimator for the binned trispectrum is $\hat{T}_{\rm opt}= \hat{T}_{\rm tot.}-\hat{T}_{\rm disc.}^{6\,\delta^2}-\hat{T}_{\rm disc.}^{\,3\langle \delta^2 \rangle^2}$, where each term can be written in terms of FFTs using Eqs. \eqref{eq:total_estimator_Tk}, ~\eqref{eq:T_disc_3_PP}, and \eqref{eq:T_disc_6_deltaP}. 

In Fig.~\ref{fig:trispectrum_disconnected_test}, we validate our trispectrum estimator using simulations of 200 Gaussian random fields with a scale-invariant power spectrum. We measure the trispectrum for all bins with $4k_f<k_1,k_2,k_3,k_4,k_{12}<10k_f$ and width $\Delta k_i=1k_f,$ where $k_f$ is the fundamental mode. The left panel shows the connected (red) and disconnected (blue) contributions to the four-point function. The trispectrum, \emph{i.e.}, the connected contribution, is consistent with zero as expected for a Gaussian random field. On the other hand, the disconnected terms are quite large and must be carefully subtracted to obtain an unbiased estimate of the trispectrum.

The right panel of Fig.~\ref{fig:trispectrum_disconnected_test} shows the ratio of the estimated trispectrum to the total four-point function using three different estimates of the disconnected contribution. The optimal estimator (black) has lower bias and variance than the naive estimators (blue and red). Since we generate the Gaussian random fields with a known fiducial power spectrum, the observed bias is solely due to numerical effects. Specifically, we simulate the Gaussian random fields with a relatively low resolution ($N_{\rm grid}=128$), thus there is a 2-3\% discrepancy between the measured power spectrum and fiducial power spectrum for the bins analyzed here. This small bias can lead to more significant biases in the trispectrum estimator, particularly at smaller scales (larger bin indices). Crucially, this bias can be partially mitigated by using the optimal estimator, which is less sensitive to errors in the fiducial power spectrum~\cite{Smith:2015uia}. We have verified that the bias disappears if we increase $N_{\rm grid}.$ Nevertheless, we choose to show the low-resolution results here to further demonstrate the utility of the optimal estimator. Indeed, realistic trispectrum analyses will likely be faced with similar percent-level inaccuracies in the fiducial power spectrum.

\end{document}